%% file: DMDCancer.tex
\documentclass[11pt]{article}

\usepackage{PRIMEarxiv}

\usepackage[utf8]{inputenc} 
\usepackage[T1]{fontenc}    
\usepackage{hyperref}       
\usepackage{url}            
\usepackage{booktabs}       
\usepackage{amsfonts}       
\usepackage{nicefrac}       
\usepackage{microtype}      
\usepackage{lipsum}
\usepackage{graphicx}
\usepackage{amsmath}
\usepackage{caption}
\usepackage{subcaption}
\usepackage{bigints}
\usepackage{multirow}
\usepackage{algorithm}
\usepackage{algorithmic}

\DeclareMathOperator*{\argmin}{arg\,min}
\usepackage{xcolor}
\usepackage[normalem]{ulem}
\newcommand{\xx}{\boldsymbol{x}}
\newcommand{\nn}{\boldsymbol{n}}

\usepackage{lineno}

\title{Data-driven simulation of Fisher-Kolmogorov tumor growth models using Dynamic Mode Decomposition
}

\author{
   Alex Viguerie \\
  Department of Mathematics\\
  Gran Sasso Science Institute\\
  Viale Francesco Crispi 7, L'Aquila, AQ 67100, Italy \\
  \texttt{alexander.viguerie@gssi.it} \\
  \And
Malú Grave \\
  Dept. of Civil Engineering\\
  COPPE/Federal University of Rio de Janeiro\\
  P.O. Box 68506, RJ 21945-970, Rio de Janeiro, Brazil \\
  Fundação Oswaldo Cruz – Fiocruz \\
  Rua Waldemar Falcão 121, BA 40296-710, Salvador, Brazil \\
  \texttt{malugrave@nacad.ufrj.br} \\
   \And
    Gabriel F. Barros \\
  Dept. of Civil Engineering\\
  COPPE/Federal University of Rio de Janeiro \\
  P.O. Box 68506, RJ 21945-970, Rio de Janeiro, Brazil \\
  \texttt{gabriel.barros@coc.ufrj.br} \\
     \And
 Guillermo Lorenzo \\
  Oden Institute for Computational Engineering and Sciences\\
  The University of Texas at Austin \\
201 E. 24th Street, Austin, TX, 78712-1229, USA \\
Dipartimento di Ingegneria Civile ed Architettura\\
  Università di Pavia \\
  Via Ferrata 3, Pavia, PV 27100, Italy \\
  \texttt{guillermo.lorenzo@unipv.it} \\
   \And
 Alessandro Reali \\
  Dipartimento di Ingegneria Civile ed Architettura\\
  Università di Pavia \\
  Via Ferrata 3, Pavia, PV 27100, Italy \\
  \texttt{alereali@unipv.it} \\
  \And
 Alvaro L.G.A. Coutinho \\
  Dept. of Civil Engineering\\
  COPPE/Federal University of Rio de Janeiro \\
  P.O. Box 68506, RJ 21945-970, Rio de Janeiro, Brazil \\
  \texttt{alvaro@nacad.ufrj.br} \\
}

\begin{document}
\maketitle

\begin{abstract}
The computer simulation of organ-scale biomechanistic models of cancer personalized via routinely collected clinical and imaging data enables to obtain patient-specific predictions of tumor growth and treatment response over the anatomy of the patient’s affected organ.
These patient-specific computational forecasts have been regarded as a promising approach to personalize the clinical management of cancer and derive optimal treatment plans for individual patients, which constitute timely and critical needs in clinical oncology.
However, the computer simulation of the underlying spatiotemporal models can entail a prohibitive computational cost, which constitutes a barrier to the successful development of clinically-actionable computational technologies for personalized tumor forecasting.
To address this issue, here we propose to utilize Dynamic-Mode Decomposition (DMD) to construct a low-dimensional representation of cancer models and accelerate their simulation.
DMD is an unsupervised machine learning method based on the singular value decomposition that has proven useful in many applications as both a predictive and a diagnostic tool.
We show that DMD may be applied to Fisher-Kolmogorov models, which constitute an established formulation to represent untreated solid tumor growth that can further accommodate other relevant cancer phenomena (e.g., therapeutic effects, mechanical deformation).
Our results show that a DMD implementation of this model over a clinically-relevant parameter space can yield impressive predictions, with short to medium-term errors remaining under 1\% and long-term errors remaining under 20\%, despite very short training periods. In particular, we have found that, for moderate to high tumor cell diffusivity and low to moderate tumor cell proliferation rate, DMD reconstructions provide accurate, bounded-error reconstructions for all tested training periods. We posit that this data-driven approach has the potential to greatly reduce the computational overhead of personalized simulations of cancer models, thereby facilitating tumor forecasting, parameter identification, uncertainty quantification, and treatment optimization.

\end{abstract}


\input{sections/1_introduction}
\input{sections/2_Methods}
\input{sections/3_Results}
\input{sections/4_Discussion}

\section*{Acknowledgments}
This research was financed in part by the Coordena\c{c}\~ao de Aperfei\c{c}oamento de Pessoal de N\'ivel Superior - Brasil (CAPES) - Finance Code 001. This research has also received funding from CNPq and FAPERJ. A. Reali was partially supported  by the Italian Ministry of University and Research (MIUR) through the PRIN project XFAST-SIMS (No. 20173C478N). G. Lorenzo acknowledges funding from the European Union's Horizon 2020 research and innovation program under the Marie Sk\l{}odowska-Curie grant agreement No. 838786.

\input{sections/5_Appendix}

\bibliographystyle{abbrv}  
\bibliography{references}

\end{document}

%% file: sections/1_introduction.tex
\section{Introduction}
\label{sec:introduction}
Cancer constitutes a global health burden: approximately 1 in 5 persons will develop cancer in their lifetime and about 1 in 10 persons will die from the disease worldwide \cite{Ferlay2021}. 
Despite the continuous advances in diagnostic and therapeutic methods, the current clinical management of cancers largely relies on historical population statistics that guide decision-making upon the observation of disease status and treatment outcome in individual patients (see, e.g., \cite{Litwin2017,Waks2019,Omuro2013}).
This approach only offers a limited personalization of cancer care and largely ignores the intrinsic high heterogeneity of cancerous diseases both within and between patients, which may result in treatment failure \cite{Marusyk2010,Jamal2015}.
Additionally, the observational nature of the current standard-of-care in clinical oncology does not enable to anticipate patient-specific prognosis, which would contribute to deliver optimal treatments maximizing outcomes and minimizing side effects for each individual patient.

Computational oncology aims at advancing the management of cancers from the current population-based observational standard to a patient-specific predictive paradigm by leveraging personalized cancer forecasts obtained via computer simulations of mathematical models that describe the key biophysical mechanisms underlying cancer phenomena \cite{Lorenzo2021,Rockne2019,Karolak2018}.
To facilitate the clinical use of this approach, the personalization of the forecasts usually relies on routinely collected patient-specific data (e.g., medical imaging, blood tests, biopsies), which enable to identify the model parameters, construct a virtual anatomic representation of the host organ and tumor, and validate model predictions \cite{Lorenzo2021,Kazerouni2020,Yankeelov2013}.
Computational oncology has been experiencing rapid growth and recent efforts have shown promise in predicting pathological and therapeutic outcomes for individual patients suffering from multiple types of cancers, such as brain \cite{Hormuth2021,Lipkova2019,Mang2020,AyalaHernandez2021,Wang2009}, breast \cite{Jarrett2020,Vavourakis2016}, prostate \cite{Colli2021,Lorenzo2016,Lorenzo2019,BradyNicholls2020}, pancreas \cite{Wong2017}, and kidney tumors \cite{Chen2013}.
Ultimately, validated models could also be leveraged to rigorously derive optimal treatment plans for individual patients \cite{Colli2021,Jarrett2020,Lorenzo2021}.

The mathematical models used to describe cancer phenomena in clinical scenarios usually rely on either ordinary differential equations (ODEs) or partial differential equations (PDEs) to respectively describe the temporal or spatiotemporal mechanisms of cancer development and response to treatments \cite{Lorenzo2021,Kazerouni2020,Rockne2019,Yankeelov2013,Mang2020,Karolak2018}. 
ODE models have met a widespread use in computational oncology due to the frequent use of scalar metrics in monitoring cancer patients (e.g., tumor volume, blood biomarkers) and their minimal computational cost, which facilitates parameter estimation, uncertainty quantification, and therapy optimization \cite{BradyNicholls2020,AyalaHernandez2021,Zahid2021,Lorenzo2019a,Brueningk2021}.
However, ODE models can only provide a limited representation of the heterogeneous tumor architecture and are not able to capture key spatially-resolved mechanisms, such as cancer cell mobility, tumor-induced mechanical deformation of host tissue, and vascular delivery of cancer therapeutics, as well as tissue-scale aspects of standard clinical interventions, such as radiotherapy and surgery \cite{Lorenzo2021,Yankeelov2013,Jarrett2020,Marusyk2010,Lipkova2019,Vavourakis2016}.
PDE models can naturally overcome these limitations and they have been attracting increasing attention as anatomical and quantitative medical imaging (e.g., computerized tomography, magnetic resonance imaging, positron emission tomography) are becoming more common in the diagnosis, staging, monitoring, and treatment of cancers \cite{Jarrett2020,Hormuth2021,Lorenzo2016,Lorenzo2019,Lipkova2019,Vavourakis2016,Wang2009}.
These imaging data provide the necessary information to construct a patient-specific virtual anatomic model and characterize the patient's tumor architecture and dynamics \cite{Lorenzo2021,Yankeelov2013,Mang2020}.
However, PDE models entail a much higher computational cost than ODE formulations, which may impede the estimation of model parameters, uncertainty in model estimations, and optimal therapeutic plans in clinically-relevant times.
Thus, there is a critical need to develop computational technologies that enable an accelerated calculation of these features and, hence, facilitate the transfer of PDE models of cancer to an actionable clinical use.

Here, we propose to leverage Dynamic Mode Decomposition (DMD) \cite{Kutz2016book} to construct an accurate and efficient lower-dimensional representation of cancer models that enable rapid computer simulations. DMD is a Scientific Machine Learning (SciML) technique that extracts the most relevant dynamical structures existent in spatiotemporal data using a purely data-driven approach, with applications ranging from short-time future estimates to control, modal analysis and dimensionality reduction \cite{Kutz2016book}. DMD has been deployed in a wide range of scientific and engineering applications such as biomechanics \cite{Calmet2020}, epidemiology \cite{Proctor2015, BGVRC2021, viguerie2022coupled}, climate \cite{Kutz2016}, aeroelasticity \cite{Fonzi2020} and urban mobility \cite{Alla2020}.
In particular, our study focuses on developing a DMD implementation of diffusion-reaction models of cancer relying on the Fisher-Kolmogorov equation.
This PDE model constitutes a established paradigm to represent untreated solid tumor growth that has been further extended to accommodate other mechanisms, such as therapeutic effects and mechanical inhibition of tumor growth \cite{Lorenzo2021,Jarrett2020,Lipkova2019,Hormuth2021,Mang2020,Wong2017,Wang2009}.
Additionally, in this work we perform a simulation study to analyze the accuracy of several DMD training and reconstruction strategies over clinically-relevant parameterizations of the model and identify the regions of the parameter space where DMD can be successfully deployed.

The rest of the paper is organized as follows. In Section \ref{sec:methods}, we describe the Fisher-Kolmogorov model in the context of tumor growth, introduce the DMD implementation strategy for this model, and outline the computational setup for the simulation studies in this work. Then, in Section \ref{sec:results} we present and analyze our numerical results. Finally, we draw conclusions and identify future lines of research in Section \ref{sec:discussion}.

%% file: sections/2_Methods.tex
\section{Methods}\label{sec:methods}
\subsection{Mathematical model}\label{sec:FisherKolmogorov}

The Fisher-Kolmogorov model consists of a second-order reaction-diffusion PDE that describes the propagation of wavefronts in a nonlinear system (see Eq.~\eqref{eqn1} below). 
This model has been applied in a myriad of different settings, including population dynamics \cite{MurrayI, MurrayII, artiles2008patch, el2019revisiting, giometto2014emerging}, epidemiology \cite{keller2013numerical, beneduci2021unifying}, and propagation of domain walls in liquid crystals \cite{guozhen1982experiments}, among others. 
In the present work, we are most interested in its application to the modeling of untreated solid tumor growth \cite{MurrayII, Lorenzo2021, Lipkova2019, Jarrett2020, Hormuth2021, Mang2020,Wong2017,Wang2009}.
In this context, the Fisher-Kolmogorov equation describes the spatiotemporal dynamics of the tumor cell density $\phi(\xx,t)$ (cells/mm$^3$) as a combination of two driving mechanisms: the tumor cell mobility, which is represented by the diffusion operator, and the tumor cell net proliferation, which is modeled via the nonlinear logistic reaction term.
Denoting the spatial domain by  $\Omega \in \mathbb{R}^{n_d}$, $n_d=1,\,2,\,3$ and the time horizon by $T$, the usual formulation of the model reads as
\begin{alignat}{2}\label{eqn1}
\frac{\partial \phi}{\partial t} &= \nabla \cdot \left(D \nabla \phi\right) + \rho \phi\left(1-\frac{\phi}{\theta}\right) \qquad &\text{ in } \Omega \times \lbrack 0,\,T\rbrack, \\ \label{eqn2}
\nabla \phi \cdot \nn &= 0 \qquad &\text{ on } \partial \Omega \times \lbrack 0,\,T\rbrack, \\ \label{eqn3}
\phi(0,\xx) &= \phi_0 (\xx) \qquad &\text{ at } t=0.
\end{alignat}
In Eq.~\eqref{eqn1}, $D$ denotes the \textit{diffusion coefficient} (mm$^{n_d}$/days), $\rho$ represents the \textit{net tumor cell proliferation rate} (days$^{-1}$), and $\theta$ is the \textit{tissue carrying capacity} (i.e., the maximum admissible tumor cell density). 
In general, these parameters may be defined pointwise and over time (i.e., $D=D(\xx,t)$, $\rho=\rho(\xx,t)$, and $\theta=\theta(\xx,t)$), for example, to account for preferential directions of growth, mechanical inhibition of tumor cell mobility and proliferation, the existence of a vascular network supporting tumor development, as well as the local and transient effect of therapies on the mechanisms in the model \cite{MurrayII, Lorenzo2021, Lipkova2019, Jarrett2020, Hormuth2021, Mang2020, Wong2017, Wang2009}.
However, we will adopt constant values for $D$, $\rho$, and $\theta$ for the sake of simplicity in this initial study on DMD reconstruction of the Fisher-Kolmogorov solutions for tumor forecasting applications.
Additionally, we will henceforth set $\theta=1$ in Eq.~\eqref{eqn1}, which is equivalent to assume that $\phi(\xx,t)$ represents the normalized tumor cell density (i.e., such that $0 \leq \phi(\xx,t) \leq 1 $).
Finally, Eq.~\eqref{eqn2} introduces no-flux boundary conditions, assuming that the tumor grows confined within the considered tissue or organ domain defined by $\Omega$ ($\nn$ is the outward unit vector normal to $\partial\Omega$) , while Eq.~\eqref{eqn3} defines the initial conditions.

Eq.~\eqref{eqn1}-\eqref{eqn3} constitute the strong form of the organ-confined untreated solid tumor growth problem.
By choosing a test function $v$ in a suitable function space $\mathcal{V}$, we obtain the variational form of \eqref{eqn1}-\eqref{eqn3} as follows: Find $\phi \in \mathcal{V}$ such that, for all $v \in \mathcal{V}$,
\begin{align}\label{variationalForm}
    \int_{\Omega} \frac{\partial \phi}{\partial t} v \, d\Omega + \int_{\Omega} D \nabla \phi \cdot \nabla v \, d\Omega - \int_{\Omega} \rho \phi\left(1-\frac{\phi}{\theta}\right) v \, d\Omega &= 0,
\end{align}
where the Neumann boundary condition \eqref{eqn2} makes the boundary integral term arising from the integration-by-parts to vanish. 
The variational problem \eqref{variationalForm} is known to admit a unique solution $\phi$, which tends to either the trivial steady state of 0 or a non-trivial, non-negative steady state as $t \to \infty$ \cite{henry2006geometric, aronson1978multidimensional}. Furthermore, provided $0\leq \phi_0 \leq 1$, then $0\leq \phi \leq 1$  for all $t$ \cite{aronson1978multidimensional}.

\subsection{Dynamic Mode Decomposition}

\par 
DMD is an equation-free data-driven method that extracts the most dynamically relevant structures on data containing spatial and temporal structures based on little to no assumptions on data. These structures are often described as DMD modes and are associated with frequency, and amplification/damping terms \cite{Kutz2016book}. It is known that the considered dataset, even if containing data from nonlinear phenomena, can be reconstructed from a significant smaller portion of DMD modes without compromising accuracy. For instance, a turbulent and highly nonlinear flow containing thousands of DMD modes can be properly reconstructed using only the first one hundred DMD modes \cite{BGVRC2021}. Besides of efficiently reconstructing accurate solutions with a significantly smaller quantity of information, DMD has been applied on a wide range of applications, from model order reduction \cite{Alla2017}, to control \cite{Fonzi2020}, temporal inference \cite{BGVRC2021}, and modal analysis \cite{Taira2017}. Despite of several variations for specific applications, herein we make use of standard DMD, which we briefly describe below.
\par 
For this study, we consider the dataset generated by the numerical simulation of the Fisher-Kolmogorov model described on Section 2.3. Each snapshot $\mathbf{x}_i$ - the solution vector obtained each time step $i$ from the numerical simulation - is stacked vertically on a matrix of dimensions $n \times (m + 1)$, where $n$ is the number of degrees of freedom of the problem and $m + 1$ is the number of time steps for which the solution is computed. This matrix, $\mathbf{X}$, is called Snapshot Matrix. The next step is to split $\mathbf{X}$ into $\mathbf{X}_1$ and $\mathbf{X}_2$ such that $\mathbf{X}_1 = [\mathbf{x}_0 \dotsc \mathbf{x}_m]  \in \mathbb{R}^{n \times m}$ and $\mathbf{X}_2 = [\mathbf{x}_1 \dotsc \mathbf{x}_{m+1}]  \in \mathbb{R}^{n \times m}$. The two matrices can be linearly mapped as
\begin{equation}
    \mathbf{X}_2 = \mathbf{A}\mathbf{X}_1,
\end{equation}
where $\mathbf{A}$ maps the dynamics of the evolution in time of the data. Although not being computationally efficient, the computation of $\mathbf{A}$ could be done directly by multiplying $\mathbf{X}_2$ to the Moore-Penrose pseudoinverse of $\mathbf{X}_1$. This approach is described as the exact DMD and can be formulated as the optimization problem
\begin{equation}
    \mathbf{A} = \argmin_{\mathbf{A}} || \mathbf{X}_2 - \mathbf{A}\mathbf{X}_1||_F
\end{equation}
\noindent where $||\cdot||$ is the Frobenius norm. Instead, the pseudoinverse can be approximated by applying the Singular Value Decomposition (SVD) factorization into $\mathbf{X}_1$ and preserving the first $r$ singular values and vectors, where $r$ must be carefully chosen to preserve the structures of the matrix. That is:
\begin{equation}
    \mathbf{X}_1  = \mathbf{U}\mathbf{\Sigma}\mathbf{V}^T,
\end{equation}
where $\mathbf{\Sigma}$ are the singular values and $\mathbf{U}$ and $\mathbf{V}$ are the left and right singular vectors, respectively. This approach dramatically reduces the algorithmic complexity of this operation while improving the conditioning of the resulting matrix. However, due to this approximation, the computation of $\mathbf{A}$ is replaced by the computation of $\mathbf{\Tilde{A}}$, which is a $r \times r$ projection of $\mathbf{A}$. From this point, the dynamics of the system can be interpreted from the eigenvectors and eigenvalues of $\mathbf{\Tilde{A}}$. The DMD modes can be obtained as:
\begin{center}
	\begin{equation}
	\mathbf{\Psi} = \mathbf{X}_2\mathbf{V}_r\mathbf{\Sigma}^{-1}_r\mathbf{W},
	\end{equation}
\end{center}
where $\mathbf{W}$ are the eigenvectors of $\mathbf{\tilde{A}}$ and the subscript $r$ indicates that $\mathbf{V}$ has been truncated to the first $r$ vectors. Then, the solution of the Fisher-Kolmogorov equation can be reconstructed and extrapolated as:
\begin{equation}\label{recon}
    \mathbf{x}(t)  \approx  \tilde{\mathbf{x}}(t) =  \mathbf{\Psi}\exp(\mathbf{\Omega}_{eig}t) \mathbf{b},
\end{equation}
with $\mathbf{b}$ being the vector containing the projected initial conditions such that $\mathbf{b} = \mathbf{\Psi^{\dagger}}\mathbf{x}_0$, and $\mathbf{\Omega}_{eig}$ is a diagonal matrix whose entries are the continuous eigenvalues $\omega_i = \ln(\lambda_i)/\Delta t_o$, where $\Delta t_o$ is the time step size between the DMD snapshots.

\subsection{Computational study setup}\label{sec_compsetup}

In this work, we perform a computational study to analyze the numerical performance of our DMD implementation in 2D and 3D test scenarios leveraging clinically-relevant parameter values from the literature \cite{Wang2009,Lorenzo2019a,Lorenzo2019,Lipkova2019,Jarrett2018,Colli2021}.
The objectives of the 2D simulation study are to evaluate the ability of DMD to learn and reproduce the dynamics of the Fisher-Kolmogorov tumor growth model, to investigate the potential relationship between the quality of the DMD reconstruction and the model parameters (i.e., $D$ and $\rho$), and to determine whether DMD shows an acceptable performance for any parameter choice or a subregion of the parameter space (i.e., a subset of tumor types).
Then, we proceed to assess whether 3D DMD reconstructions obtained for parameter values in this subregion also exhibit a satisfactory performance, which would ultimately support the use of our approach in patient-specific, organ-scale scenarios.

\begin{figure}[ht!]
\centering
\includegraphics[width=.42\linewidth]{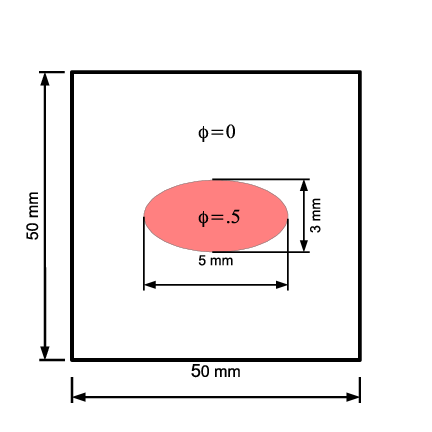}\includegraphics[width=.48\linewidth]{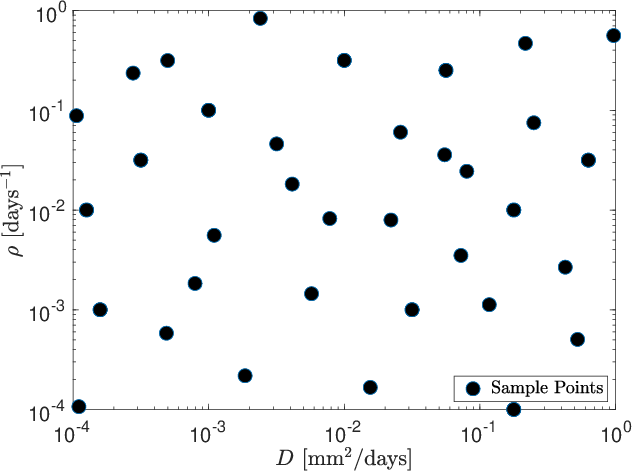} \caption{Left: Problem setup for 2D test cases: we consider a 50 mm$\times$ 50 mm tumor domain, with initial conditions defined as a tangent hill, such that $\phi$=0.5 inside the tumor and 0 outside. Right: Values of diffusion rate $D$ and proliferation rate $\rho$ considered in the 2D simulation study. Values were generated using logarithmic Latin-hypercube sampling.  }\label{fig:SamplePts}
\end{figure}

\begin{figure}[ht!]
     \centering

         \includegraphics[width=.67\textwidth]{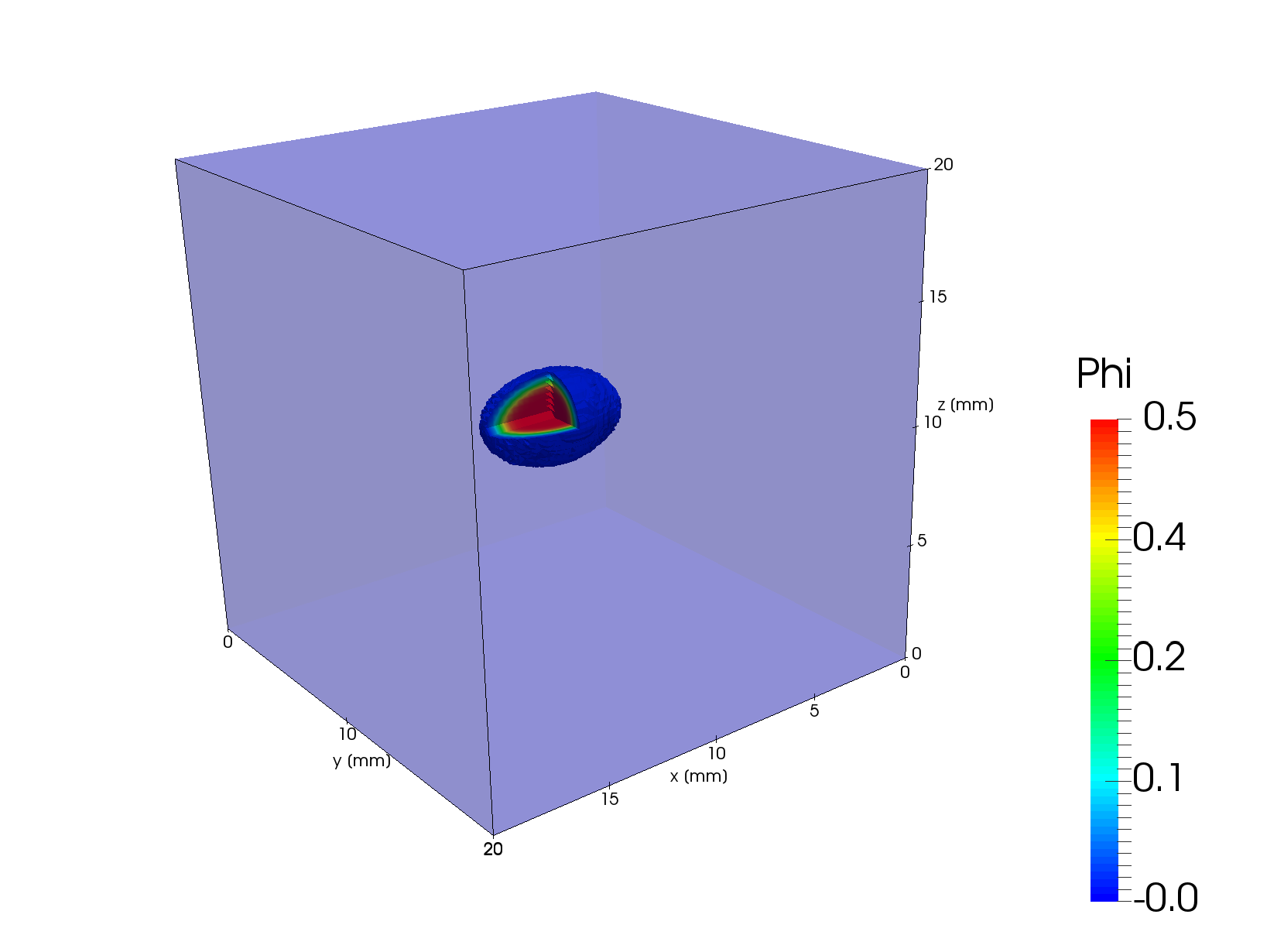}
         \caption{Problem configuration, 3D test case}\label{fig:3DProb}
\end{figure}

In 2D, we consider an elliptical tumor with diameter lengths of 3 and 5 millimeters that is centered in a square tissue domain of 50$\times$50 millimeters, as shown in Fig.~\ref{fig:SamplePts}.
Similarly, in 3D, we study an ellipsoidal tumor with diameter lengths of 5, 3, and  3 millimeters that is centered in a cubic tissue domain of 20$\times$20$\times$20 millimeters (Figure \ref{fig:3DProb}). We construct these initial conditions using a hyperbolic tangent hill function that yields $\phi=0.5$ inside the tumor and $\phi=0$ outside:
\begin{equation}\label{eq:ini2d}
    \phi(\xx,0) = \frac{1}{4} - \frac{1}{4}\tanh\left( \frac{100}{\sqrt{2}}\left( \sqrt{ \sum_{i=1}^{n_d} \frac{\left( x_i -c_i\right)^2}{d_i^2} }  - 1\right) \right),
\end{equation}
 where $\xx=\{x_i\}_{i=1}^{n_d}$ is the position vector,  $\mathbf{c}=\{c_i\}_{i=1}^{n_d}$ denotes the center of the domain, and $\{d_i\}_{i=1}^{n_d}$ are the principal diameters of the tumor initial geometry in $\mathbb{R}^{n_d}$. 
 
We construct the parameter space by leveraging values of $D$ from 10$^{-4}$ to 1 mm$^2$/days and values of $\rho$  from 10$^{-4}$ to 1 days$^{-1}$.
This broad parameter ranges include a wide array of clinically-relevant scenarios as reported in previous computational oncology studies  \cite{Wang2009,Lipkova2019,Jarrett2018}.
To explore the performance of our DMD implementation over this parameter space, we sparsely and randomly sample 35 parameter value pairs $\{D,\rho\}$ using Latin hypercube sampling, as shown in Fig.~\ref{fig:SamplePts} (for numerical data, see Appendix). For the 3D simulation, we let $D=0.00186$ mm$^3$/days, and $\rho=0.00022$ days$^{-1}$, which correspond to those used in case 23 in the 2D study. We note that the units of $D$ are now changed to reflect the shift to three spatial dimensions.

The reference simulations of the model are carried out using the  \texttt{libMesh} finite element library \cite{libmesh}.
In 2D, we discretize the computational domain using 125,000 linear triangular elements. In 3D, we use an adaptive mesh refinement/coarsening scheme (AMR/C). The mesh has initially $10 \times 10 \times 10$ cells, with each cell divided into 6 linear tetrahedra. We refine the initial region where the tumor is located into four levels and, after the refinement, the smallest element has a size of 0.125 mm. The adaptive mesh refinement is based on the flux jump of the of tumor cell density function error, in which the maximum level of refinement is equal to four. We apply the AMR/C every 20 time steps. Details about the AMR/C procedure may be found in \cite{rossa2013parallel, grave2020new}.  We output the projected solution to a fixed grid to enable the use of DMD \cite{BGVRC2021}. The projected mesh comprises 288,597 linear tetrahedra elements, with an element size of 0.25 mm. 
We integrate in time using the second order backward differentiation formula (BDF2) scheme with a constant time step of 0.25 days  \cite{Jarrett2018,Jarrett2020}. 
The resulting linear systems are solved utilizing the GMRES algorithm with a ILU(0) preconditioner. For both the 2D and the 3D case, we output the solution every 4 time steps.


The DMD method is implemented by leveraging the PADMe\footnote{https://github.com/gf-barros/padme} library \cite{Ramses2021}.
We consider 20 modes in our DMD reconstructions, since we empirically observed that they are typically enough to yield sufficient reconstruction capability after preliminary sensitivity analyses of the SVD  (see, e.g., \cite{BGVRC2021}). 
Ideally, we would like DMD computations to provide reasonable approximations for long-term model predictions with minimal-to-moderate training. 
In such a case, DMD would enable the rapid evaluation of different tumor growth scenarios by requiring the resolution of only a relatively small number of time steps. 
Thus, in order to evaluate the efficacy of DMD, we compare the performance achieved when leveraging training periods of 20, 50, 100, and 200 days within a fixed time horizon of 365 days. 
To this end,  we evaluate the relative error in $L^2$ norm $e_{DMD}^n(t)$ at a given instance $t$ as:
\begin{align}\label{Error}
e_{DMD}^{n}(t) &= \sqrt {\frac{\displaystyle \int_{\Omega} (\phi_{r}(t) - \phi_{DMD}^n (t) )^2 d\Omega     }{ \displaystyle \int_{\Omega} \phi_{r} (t)^2 d\Omega }}.
\end{align}
where $\phi_{r}$ denotes the value of $\phi$ computed in the reference simulation and $\phi_{DMD}^n$ represents the corresponding DMD reconstruction over $n$ training days.
We further construct a grading system to characterize the performance of DMD based on $e_{DMD}^n(t)$ over the whole time domain and at specific timepoints representing short-to-long-term predictions:
\begin{itemize}
    \item Class $A$: If a case is such that 
    $$ \frac{1}{365}\int_{0}^{365} e_{DMD}^{50} (t) dt < 0.1,\,\,\text{ and } e_{DMD}^{50}(365)<0.2, $$
    implying that a short training period provides an accurate reconstruction over the whole time interval.
    \item Class $B$: Cases not in Class $A$ such that:
    $$ e_{DMD}^{50}(100)<0.1,\,\,e_{DMD}^{100}(150)<0.1,\,\,e_{DMD}^{200}(250)<0.1,$$
    signifying cases that do not necessarily have good long-term error performance, but are accurate over the medium-to-short term.
    \item Class $C$: Cases not in Class $A$ or $B$ such that:
    $$ e_{DMD}^{200}(250)<1,$$
    meaning cases for which the DMD reconstruction is not necessarily accurate but for which the error remains bounded.
    \item Class $D$: All other cases, i.e., instances in which DMD does not produce helpful long- or short-term extrapolations, and for which the error may blow up.
\end{itemize}
We acknowledge that, like all scoring/classification systems, our grading system is to some extent arbitrary. However, we feel that our rigorous criteria provide a reasonable indicators for the qualitative error behaviors which we seek to identify.

%% file: sections/3_Results.tex
\section{Results}
\label{sec:results}

\subsection{Exploring the performance of 2D DMD reconstructions over a clinically-relevant parameter space}

\begin{figure}
\centering
\includegraphics[width=\linewidth]{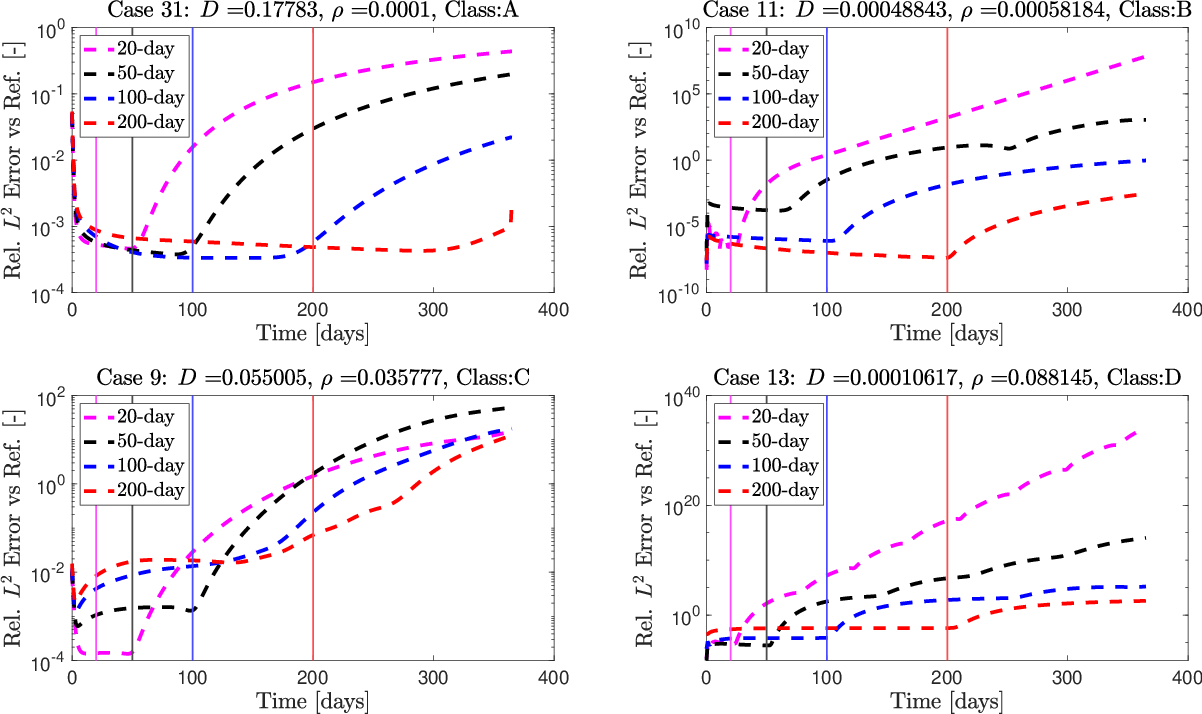} \caption{Examples of the different qualitative and quantitative classes of DMD performance according to our classification system. Legend refers to length of the training period.  }\label{fig:2dGradeExamples}
\end{figure}

\par Of our 35 cases, we obtained numerical results for 34, with the reference solution of one of the cases (case 17) failing to converge. This case has been classified as a D. In Fig. \ref{fig:2dGradeExamples}, we show the error performance of one case from each class: case 31 (A), case 11 (B), case 9 (C) and case 13 (D). These give an illustration of the general performance behavior that the classes are designed to indicate. 
Complete plots of all 34 converged cases cases are shown in Fig. \ref{fig:Case1to3}-\ref{fig:Case34to35}. Note these figures have been placed at the end of the manuscript for reasons of readability. 
We note that the vertical lines mark the end of each training period for the correspondingly-colored DMD reconstruction. We immediately note that DMD works very well for some cases, across all training levels and time intervals, and less well for others. 

\par
\begin{figure}
\centering
\includegraphics[width=.5\linewidth]{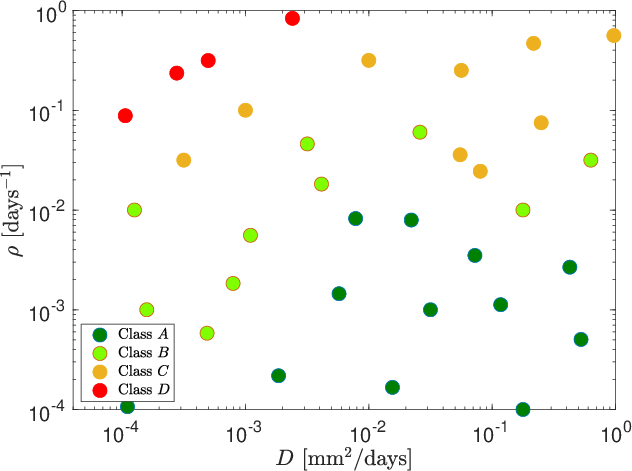} \caption{Visualization of the performance of DMD reconstructions in terms of $\rho$, $D$. The results depicted in this plot show a clear pattern in terms of our grading criteria. We can expect DMD to work well for cases in which values of $D$ are moderate to high and values of $\rho$ low to moderate.}\label{fig:2dGrades}
\end{figure}

 Applying our grading system to the results, we visualize the classes according to $\rho$ and $D$ in Fig. \ref{fig:2dGrades}. We see a clear and definite pattern to the results, and the classes form clusters in the parameter space. For moderate to high values of $D$ coupled with low to moderate values of $\rho$, we expect DMD to perform best. In such cases (Class $A$), the long-term accuracy of DMD remains high, even for short training periods. For cases in which the values of $\rho$ and $D$ are both low, we see a deterioration of the long-term error behavior, while the short-term error behavior remains nonetheless strong (Class $B$). Cases in which both $\rho$ and $D$ are high tended to yield less accurate results, while, in cases in which $\rho$ was high but $D$ was low, the DMD reconstructions tend to fail. 

\par
\begin{figure}
     \centering
     \begin{subfigure}[b]{0.31\textwidth}
         \centering
         \includegraphics[width=\textwidth]{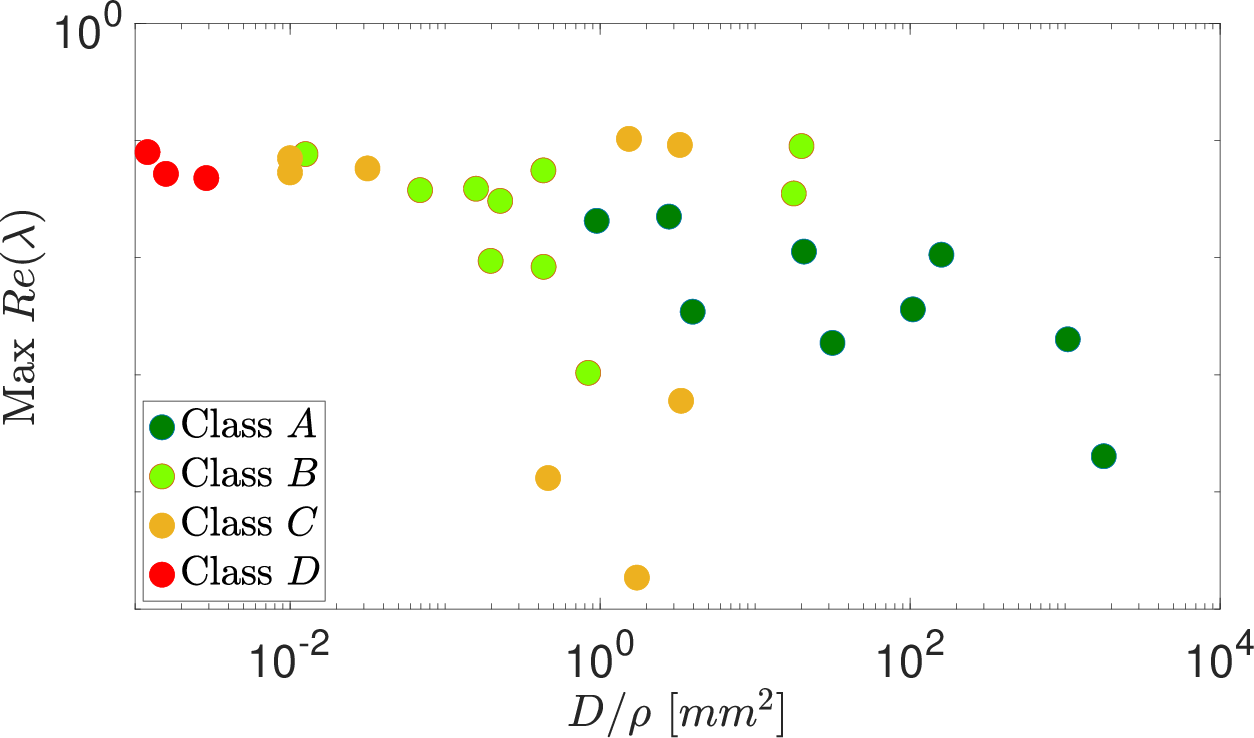}
                \caption{Relationship between max Re($\lambda$), $D/\rho$ (log scale).}

         \label{fig:ReSpectrum}
     \end{subfigure}
     \hfill
     \begin{subfigure}[b]{0.31\textwidth}
         \centering
         \includegraphics[width=\textwidth]{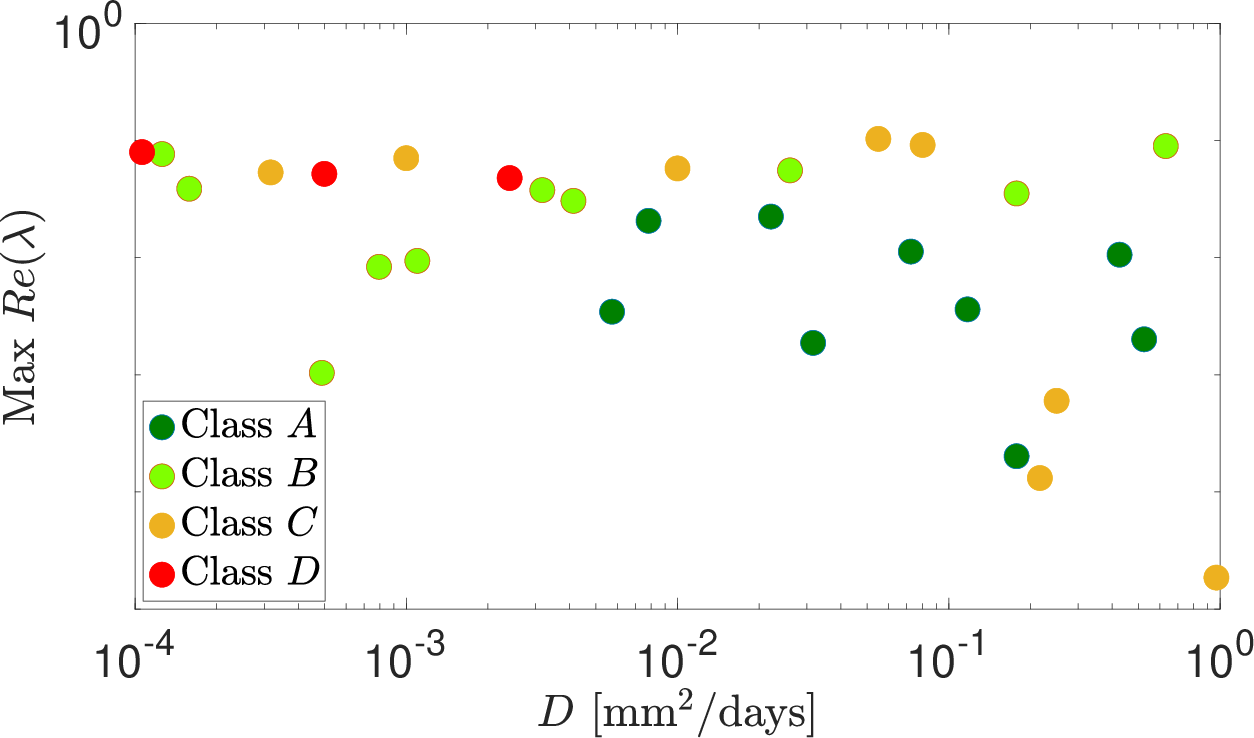}
                  \caption{Relationship between max Re($\lambda$), $D$ (log scale).}
\label{fig:NuSpectrum}
     \end{subfigure}
     \hfill
     \begin{subfigure}[b]{0.31\textwidth}
         \centering
         \includegraphics[width=\textwidth]{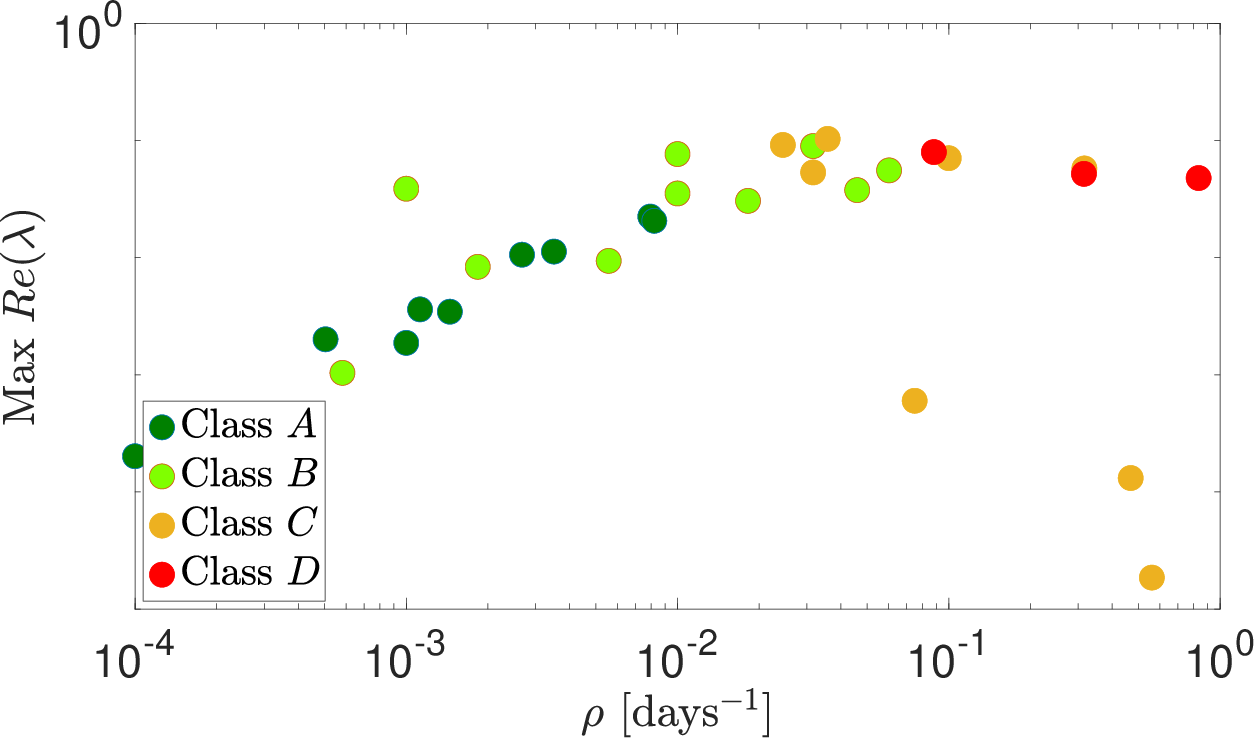}
         \caption{Relationship between max Re($\lambda$), $\rho$ (log scale).}
         \label{fig:RhoSpectrum}
     \end{subfigure}
        \caption{Max Re($\lambda$) for each DMD reconstruction and associated  $D/\rho$ (left), $D$ (center) and $\rho$ (right), colored according to the classification system. We see that the ratio $D/\rho$ and parameter $\rho$ provide strong predictors of this value, which is in turn a strong predictor of DMD performance.}
        \label{fig:SpectralAnalysis}
\end{figure}

 We then examined the influence of the maximum real part of the DMD eigenvalues for each case. Given the expression \eqref{recon}, one observes that, if some $\lambda_i$ have large real parts, we can expect the DMD performance to deteriorate, particularly for larger $t$. In Fig. \ref{fig:SpectralAnalysis}, we plot max $\text{Re}(\lambda)$ for each case, and analyze its relationship with the ratio $D/\rho$ (left), $D$ (center), and $\rho$ (right). The first thing we note is that max Re($\lambda$) is, as expected, a reliable predictor of the overall DMD performance. The higher-class cases are lower on the vertical axis representing max $\text{Re}(\lambda)$, with lower-class cases towards the higher end. In terms of relationship with the problem parameters, we see moderately strong relationships between Re($\lambda)$ and $D/\rho$, $\rho$, with the relationship less strong between Re($\lambda)$ and $D$. Quantitatively, the log-correlation coefficients are 0.574 between Re($\lambda$) and $D/\rho$, -0.208 between Re($\lambda$) and $D$, and 0.626 between Re($\lambda$) and $\rho$. 
Consequently, $D/\rho$ and $\rho$ are promising predictors of DMD performance according to max Re($\lambda$).

\par From a physical point of view, the results from our 2D parameter study shown in Figs. \ref{fig:2dGrades} and \ref{fig:SpectralAnalysis} suggest that DMD may be well-suited to reproduce tumors with low to moderate proliferation activity, which exhibit class A and B according to our grading system across the investigated range of $D$ values. For instance, this parameter subregion can represent breast cancer \cite{Jarrett2018} as well as some brain tumors \cite{Wang2009}. Conversely, our results suggest that a DMD reconstruction cannot accurately represent highly proliferative tumors irregardless of their estimated tumor cell diffusion coefficient, since the DMD reconstruction only reaches grade C at best. For example, this parameter subregion would correspond to many cases of high-grade gliomas, including glioblastoma multiforme  \cite{Lipkova2019,Wang2009}.
In particular, DMD fails to reconstruct (class D) highly-proliferative tumors with low tumor cell diffusion coefficient (i.e., with low $D/\rho$ ratio), which would correspond, for example, to nodular high-grade gliomas \cite{Baldock2014}.

\subsection{Assessing the performance of 3D DMD reconstructions in representative higher-class scenarios}

\begin{figure}[ht!]
     \centering
     \begin{subfigure}[b]{0.49\textwidth}
         \centering
         \includegraphics[width=\textwidth]{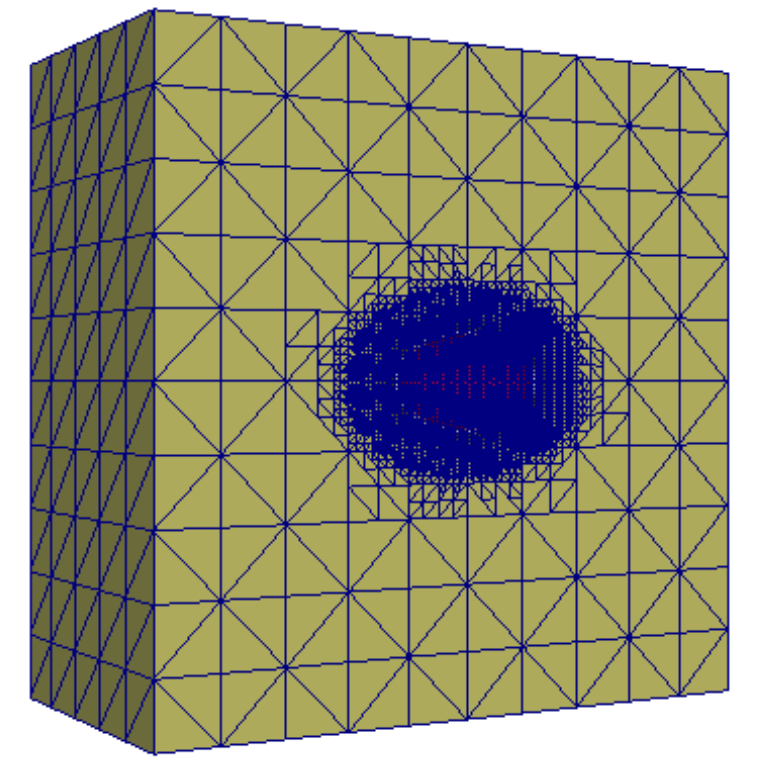}
         \caption{Adaptive mesh solution}
         \label{fig:3DCaseAdaptiveMesh}
     \end{subfigure}
     \begin{subfigure}[b]{0.49\textwidth}
         \centering
         \includegraphics[width=\textwidth]{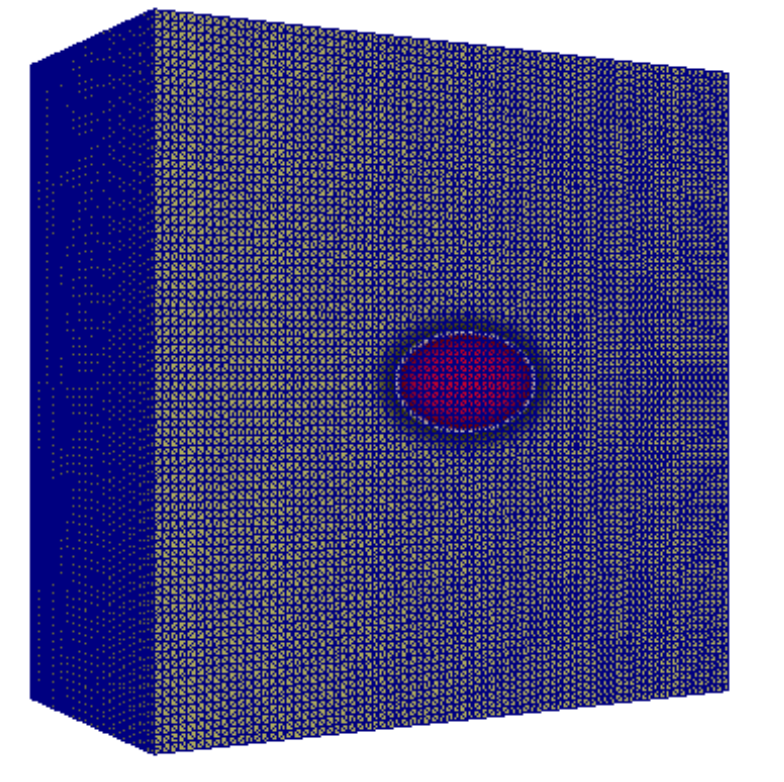}
         \caption{Mesh projection}
         \label{fig:3DCaseProjectedMesh}
     \end{subfigure}
        \caption{Solution detail at $t = 20$ days; (a) Adaptive mesh and (b) Reference mesh and projected solution. Mid-plane view.}
        \label{fig:projection}
\end{figure}

The 3D simulation correspond to those used in case 23 in the 2D study. We recall that  an important distinction from the 2D simulations is that we employ adaptive mesh refinement and coarsening during the computation; at each time step, the solutions computed on the adapted meshes are projected onto a reference mesh, sufficiently fine to resolve all relevant dynamics (see Section \ref{sec_compsetup}). This process is depicted in Figs. \ref{fig:3DCaseAdaptiveMesh} and \ref{fig:3DCaseProjectedMesh}. Further details can be found in \cite{BGVRC2021}. 

\begin{figure}[ht!]
     \centering

         \includegraphics[width=.5\textwidth]{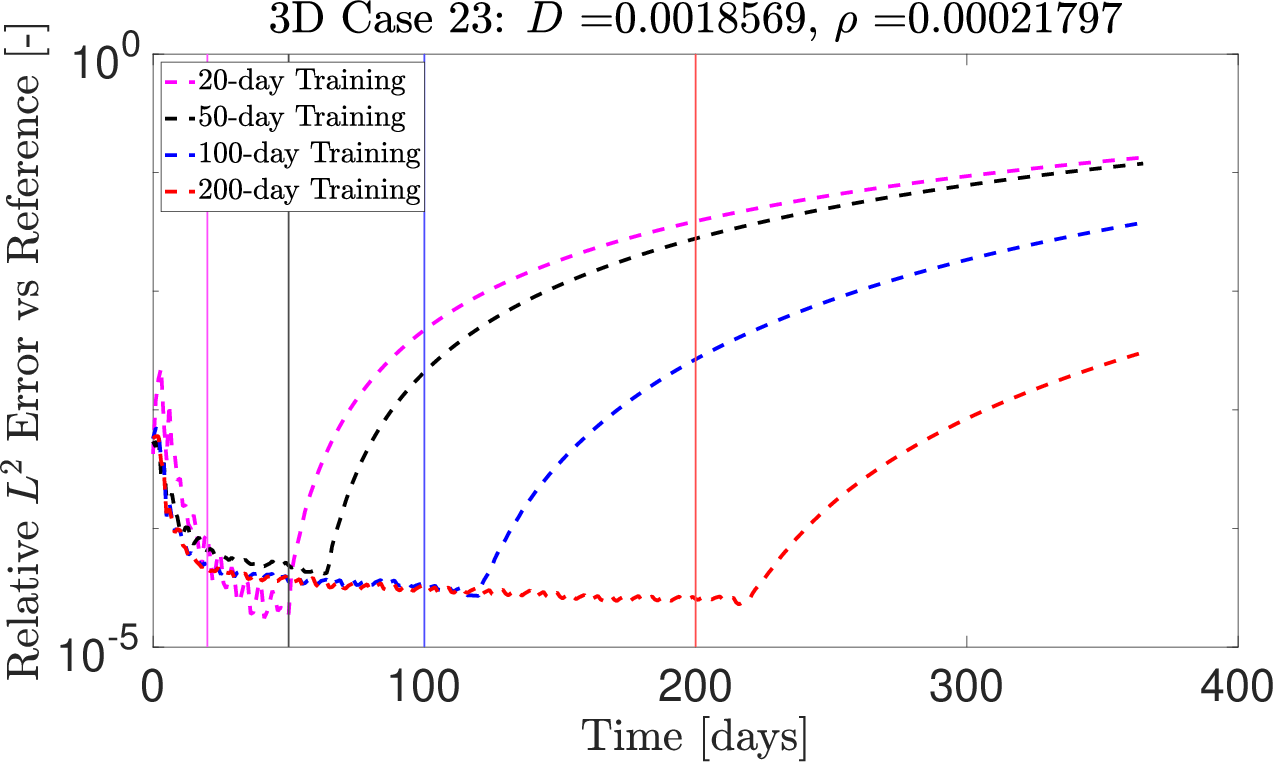}
         \caption{Results for the 3D case over the different training intervals. We see error behavior in line with the corresponding 2D behavior, suggesting that DMD may be applied similarly to three-dimensional problems}\label{fig:3Derror}
\end{figure}

\begin{figure}[ht!]
     \centering

         \includegraphics[width=\textwidth]{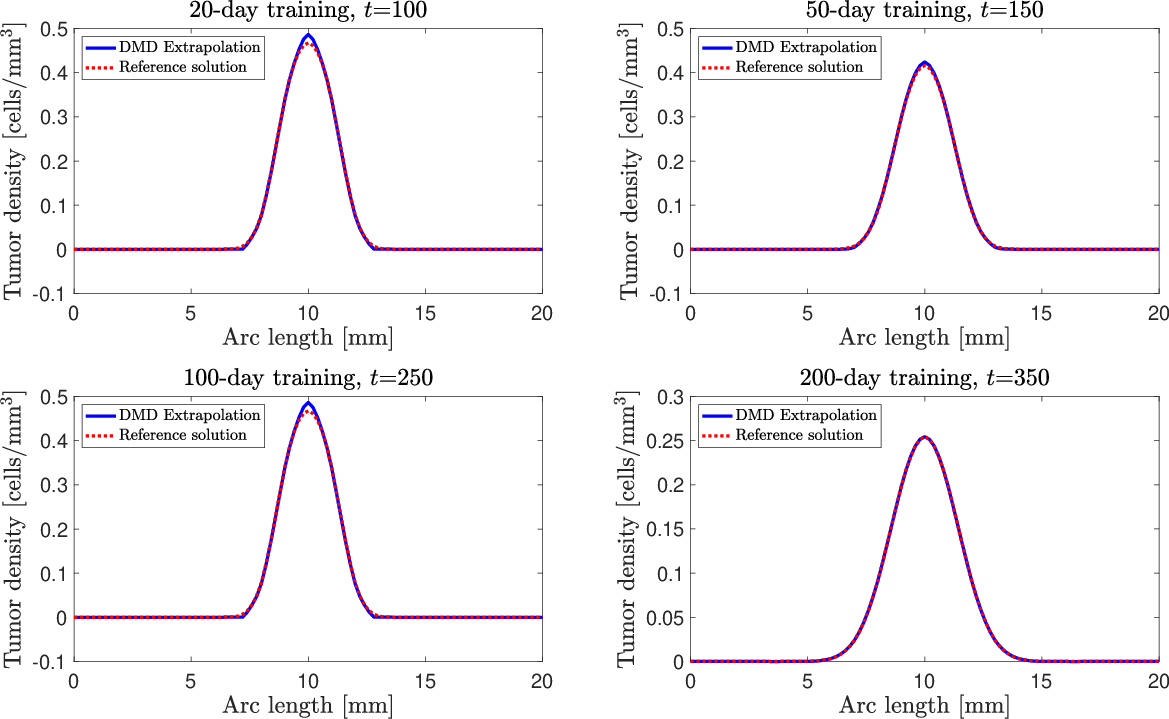}
         \caption{The tumor density plotted over the line $z=0$ to $z=20$ mm in the center of the domain. Frop top left, clockwise: $t=$100 days for 20-day training, $t=$150 days for 50 day-training, $t=$350 days for 200 day training, and $t=$250 days for 100 day training. In each case, we see good agreement with the reference solution, despite being significantly outside the training period. }\label{fig:MassCenterline}
\end{figure}

\par In Fig. \ref{fig:3Derror}, we plot the error behavior of the 3D test case. We note that it is in line with the performance of the analogous two-dimensional case (case 23). This suggests that the observed performance of DMD in two dimensions will remain similar when applied to large-scale, three-dimensional problems of this type. 
\par We display qualitative agreement of the DMD and the reference solutions in Fig. \ref{fig:MassCenterline}. We show the tumor density plotted over the line centered at $(10,10)$ and extending from $z=0$ to $z=20$ at $t=100,\,150,\,250,\,350$ for the 20-, 50-, 100-, and 200-day training periods respectively. In each case, the solution is extrapolated far beyond the conclusion of the training period; nonetheless, we observe strong agreement in each case, with the key solution characteristics reproduced successfully.
\par We note that the DMD algorithm applied to this problem required tens of seconds on a standard workstation, compared to hours for the full-order numerical simulation. Given the reasonable error performance, especially for short to medium-term, and even longer-term extrapolations, the ability to use DMD while maintaining sufficiently accurate solutions may offer a pathway for potentially large reductions in the numerical simulation cost of personalized tumor forecasts informed by routine clinical and imaging data.

\begin{figure}[ht!]
     \centering

         \includegraphics[width=.6\textwidth]{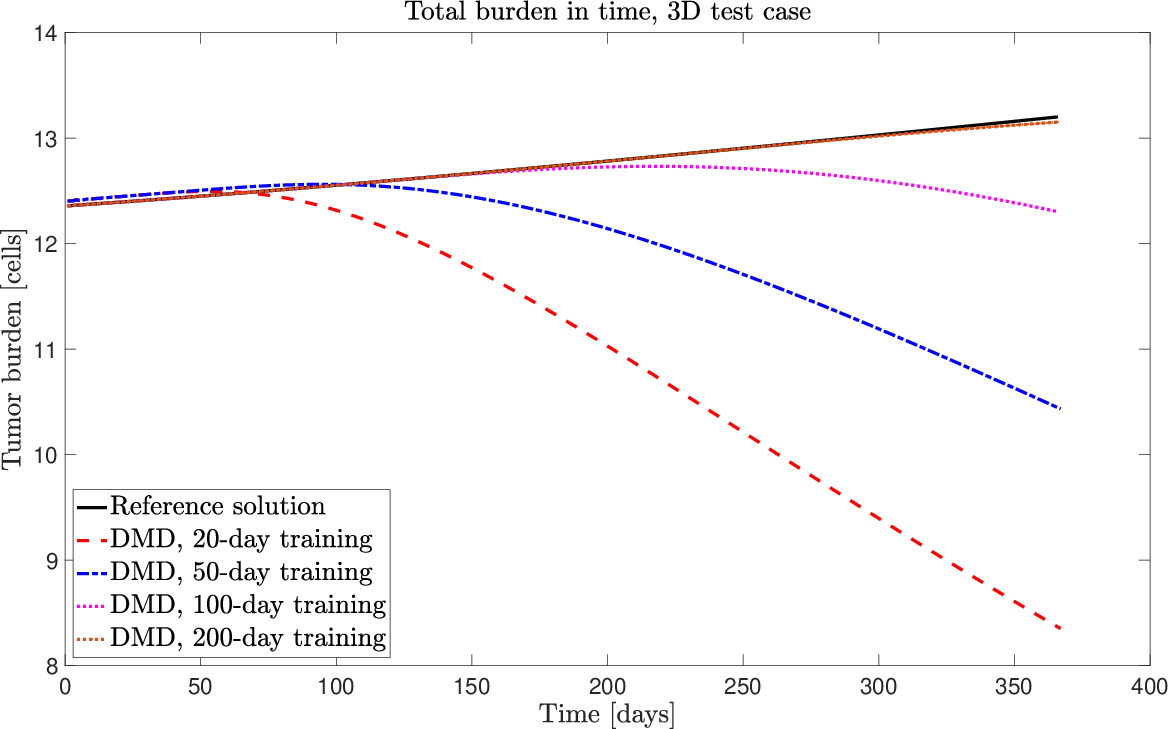}
         \caption{The tumor burden in time for the three-dimensional case. We see progressively  better agreement with the increase in training period; wit the 200-day training period being nearly indistinguishable from the reference. We also observe the 20-day, 50-day and 100-day training period solutions showing good agreement with the reference solution for long after the conclusion of the training periods.  }\label{fig:MassPlot}
\end{figure}

%% file: sections/4_Discussion.tex
\section{Discussion}
\label{sec:discussion}
In the present work, we have examined the suitability of Dynamic Mode Decomposition (DMD) applied to Fisher-Kolmogorov models for solid tumor growth. For many other problems, DMD has been shown to offer good short-to-medium term predictions, at a fraction of the cost of full-order simulations. However, its applicability is not universal, and in some instances DMD projections may be unsatisfactory. Thus, we sought to clearly identify for what classes of tumor dynamics one may apply DMD, and what level of accuracy may be expected from the reconstructions. 
\par We performed a parameter study over a sample of 35 different model parameterizations, with varying diffusion $D$ and proliferation rates $\rho$ over a range of clinically-relevant values. As expected, DMD performance ranged from accurate, even for long-term simulations, to unsatisfactory, depending on the specific parameter values. We developed a classifying system, and showed that for tumors characterized by moderate to high diffusion and low to moderate proliferation, DMD may offer good predictions over all time levels. For other types of tumor dynamics, the results are mixed; in general, DMD may still offer valuable information for all cases except for those in which the proliferation rate is high, especially if the diffusion coefficient is also low.
\par We then demonstrated the practical importance of this study by validating results on a large-scale, three-dimensional case. We found that the error behavior was in line with the two-dimensional analogue, suggesting that one may extrapolate from our two-dimensional parameter study into the corresponding three-dimensional scenarios. The DMD extrapolation for the three-dimensional simulation, for which the full-order solution requires hours, can be resolved in a matter of seconds. Indeed, the potential savings offered by this approach are so dramatic that, even if only short-term forecasts can be used reliably, DMD still offers potential computational value. For cases in which medium- and long-term projections are reliable, DMD offers a compelling alternative to full-order simulation.
\par While our results show a promising application of DMD for computational oncology applications, the study as presented has several limitations, which we plan to investigate in future studies. Firstly, we have considered only fixed, constant values for both $\rho$ and $D$; however these parameters may further depend on time, space, or even on the tumor density itself \cite{Lorenzo2021,Yankeelov2013,Mang2020}. Secondly, our studies in this work rely on numerical simulation data exclusively. Such data are inherently high-fidelity, and are not affected by the noise which may be present in clinical and imaging data from individual patients. Thus, further studies need to address to what extent noise in the input routine clinical and imaging data may affect the DMD solution reliability. Thirdly, we focused our analysis on the Fisher-Kolmogorov model, but solid tumor growth can be also represented with other spatiotemporal mathematical formulations, such as the phase-field method \cite{Lorenzo2021,Lorenzo2016,Lorenzo2019,Colli2021,Agosti2018,Lima2016}. Thus, the analysis of the performance of our DMD approach to reconstruct the solution of phase-field models of tumor growth may also be an interesting direction of work. Finally, our study focused on the reconstruction of model solution for a known parameter set. In the future, we would like to use DMD, together with projection and/or interpolation techniques, to reconstruct space-time solutions for unknown parameter values. Ideally, we may build a solution database of high-fidelity space-time solutions over a parameter space, using DMD-reconstructed solutions to evaluate unknown parameter values \cite{GAO2022110907, hess2022data}. This development could dramatically accelerate patient-specific parameter identification, Bayesian approaches in solid tumor growth modeling, and treatment optimization, which require multiple model evaluations.
Thus, the ability of DMD implementations to obtain fast model solutions for any parameter combination could ultimately facilitate a robust, rapid, and accurate calculation of patient-specific tumor forecasts to guide clinical decision-making \cite{Lorenzo2021,Yankeelov2013,Mang2020}.

%% file: sections/5_Appendix.tex
\section*{Appendix}
\label{sec:appendix}

\setcounter{equation}{0}
\setcounter{figure}{0}
\setcounter{table}{0}
\makeatletter
\renewcommand{\theequation}{A\arabic{equation}}
\renewcommand{\thefigure}{A\arabic{figure}}
\renewcommand{\thesection}{A-\arabic{section}}
\renewcommand{\thetable}{A\arabic{table}}

\par
\begin{figure}
     \centering
     \begin{subfigure}[b]{0.31\textwidth}
         \centering
         \includegraphics[width=\textwidth]{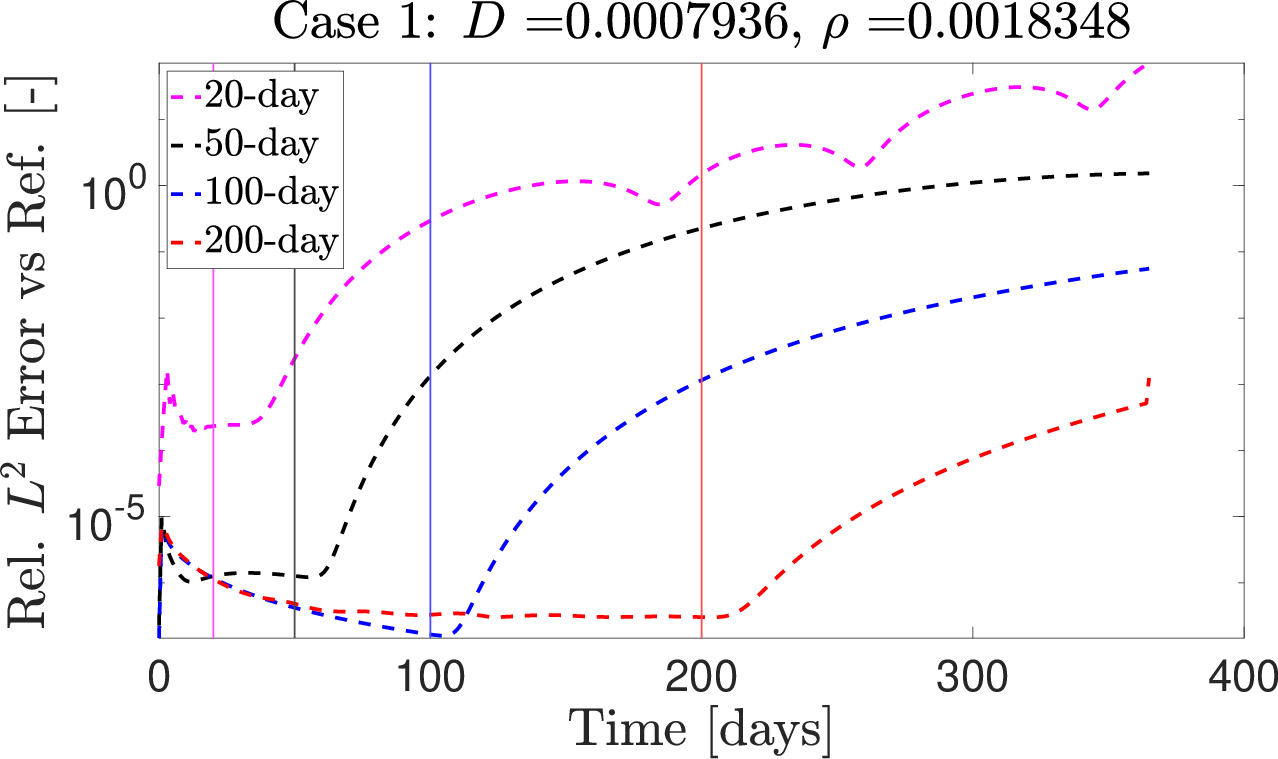}
         \caption{Case $ 1$}
         \label{fig:Case1}
     \end{subfigure}
     \hfill
     \begin{subfigure}[b]{0.31\textwidth}
         \centering
         \includegraphics[width=\textwidth]{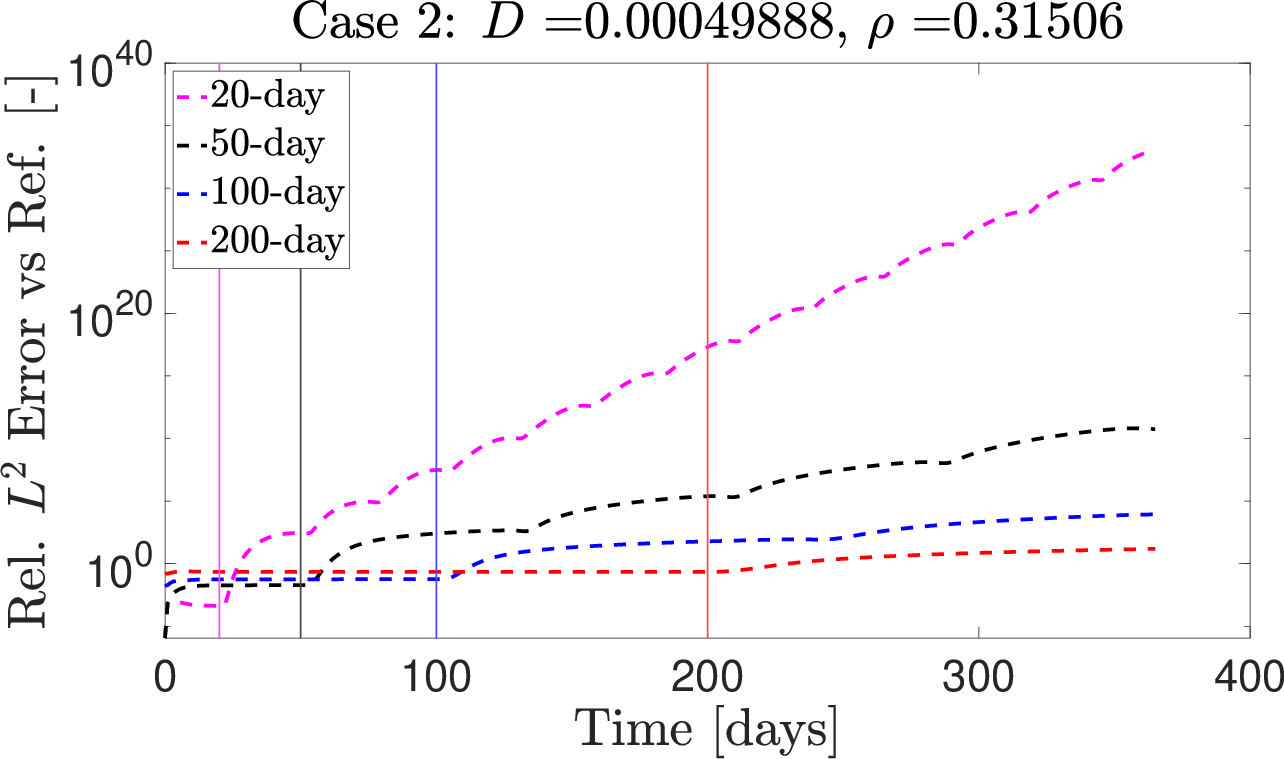}
         \caption{Case $ 2$}
         \label{fig:Case2}
     \end{subfigure}
     \hfill
     \begin{subfigure}[b]{0.31\textwidth}
         \centering
         \includegraphics[width=\textwidth]{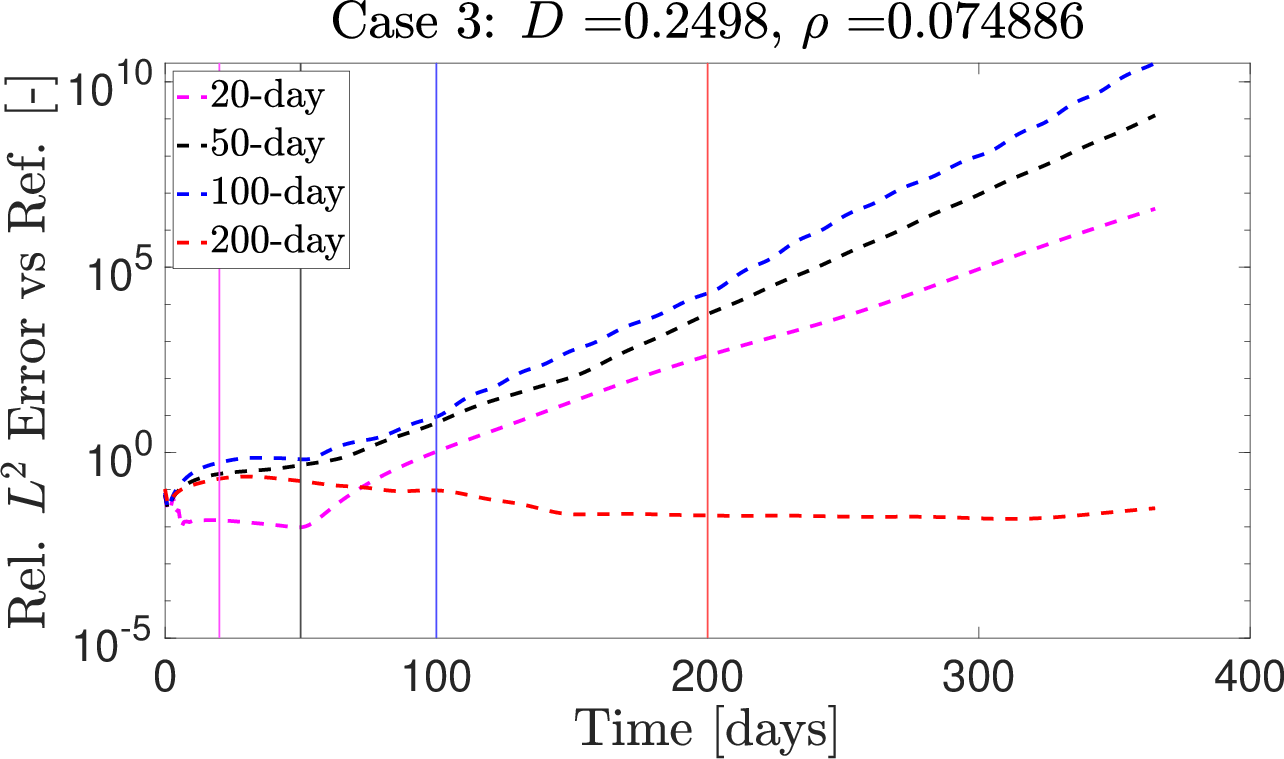}
         \caption{Case $ 3$}
         \label{fig:Case3}
     \end{subfigure}
        \caption{2D simulation study, Cases 1-3. Legend refers to length of the training period}
        \label{fig:Case1to3}
\end{figure}

\begin{figure}
     \centering
     \begin{subfigure}[b]{0.31\textwidth}
         \centering
         \includegraphics[width=\textwidth]{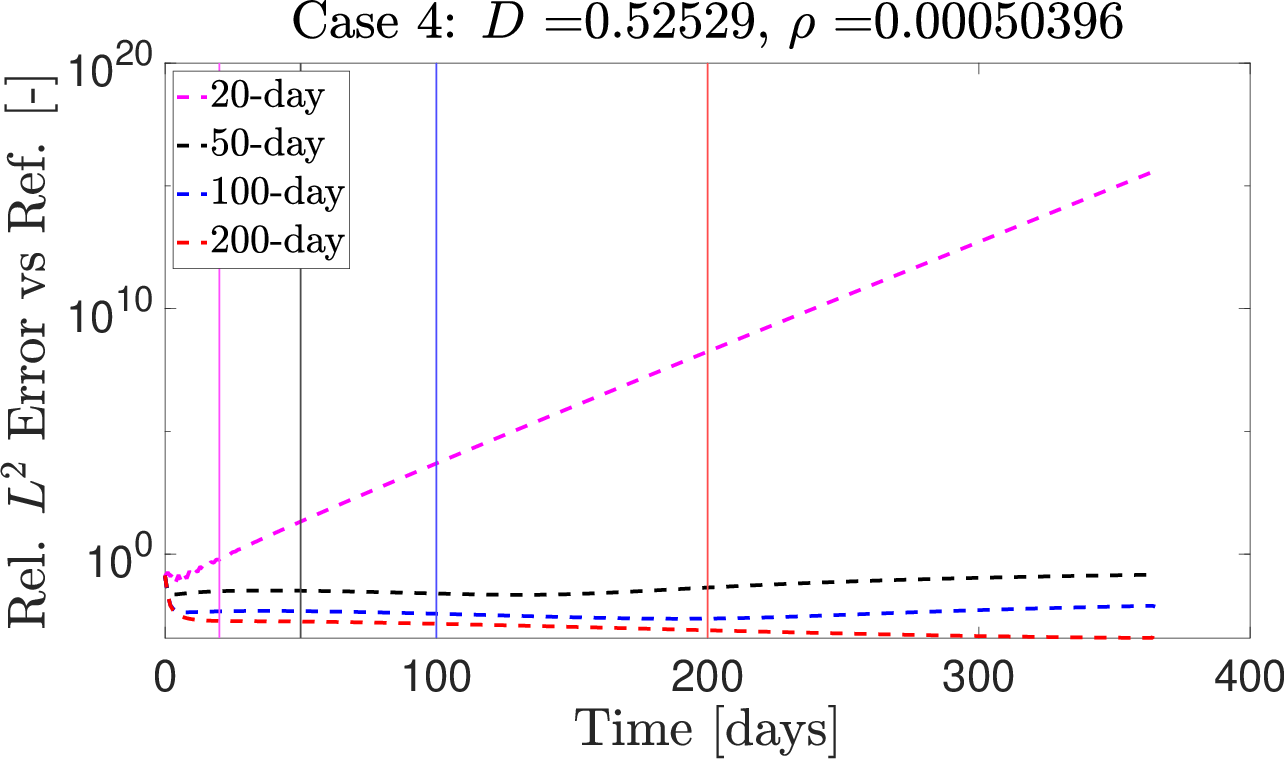}
         \caption{Case $ 4$}
         \label{fig:Case4}
     \end{subfigure}
     \hfill
     \begin{subfigure}[b]{0.31\textwidth}
         \centering
         \includegraphics[width=\textwidth]{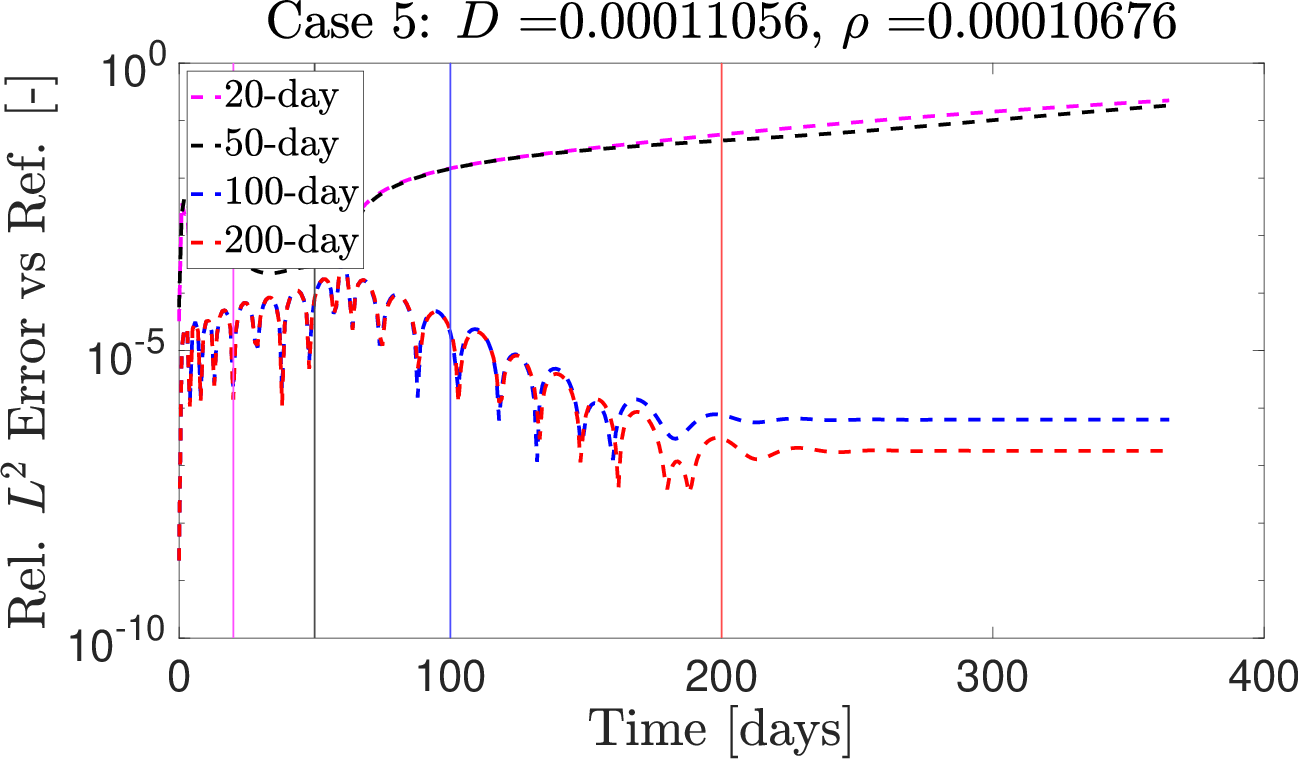}
         \caption{Case $ 5$}
         \label{fig:Case5}
     \end{subfigure}
     \hfill
     \begin{subfigure}[b]{0.31\textwidth}
         \centering
         \includegraphics[width=\textwidth]{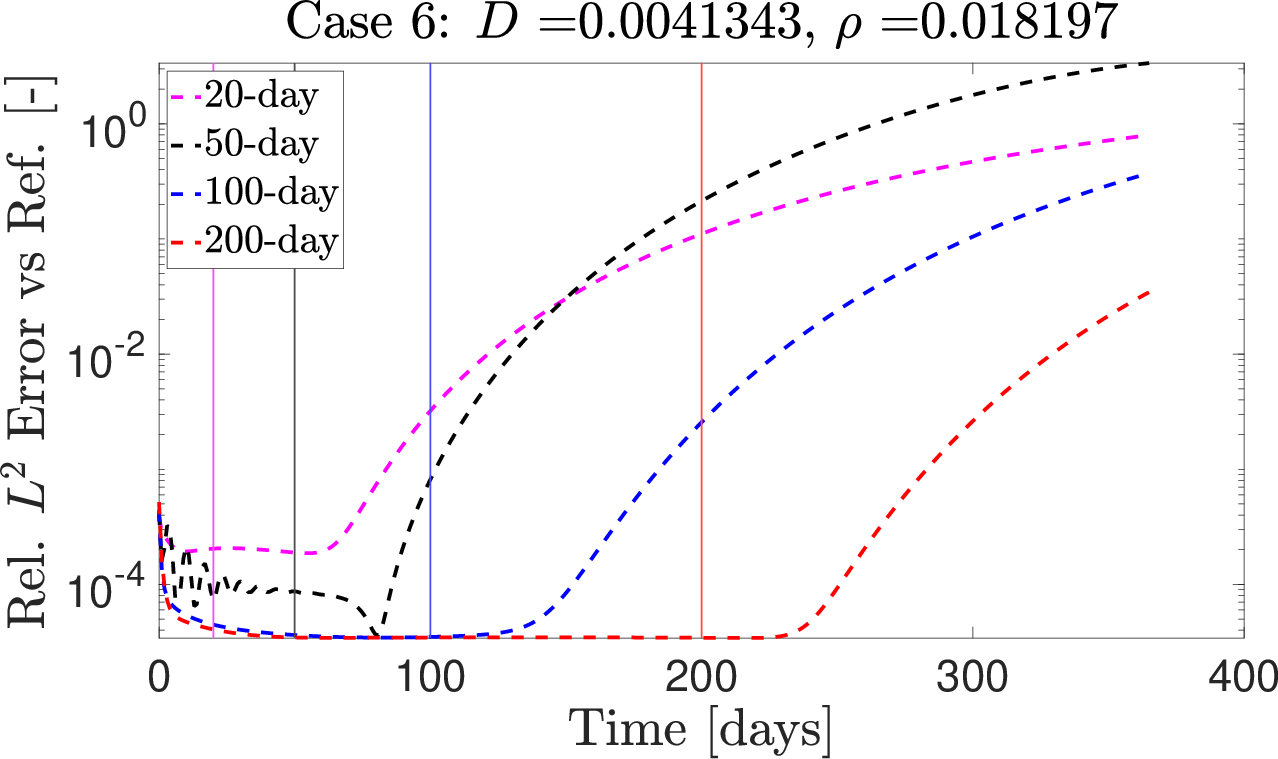}
         \caption{Case $ 6$}
         \label{fig:Case6}
     \end{subfigure}
        \caption{2D simulation study, Cases 4-6. Legend refers to length of the training period.}
        \label{fig:Case4to6}
\end{figure}

\begin{figure}
     \centering
     \begin{subfigure}[b]{0.31\textwidth}
         \centering
         \includegraphics[width=\textwidth]{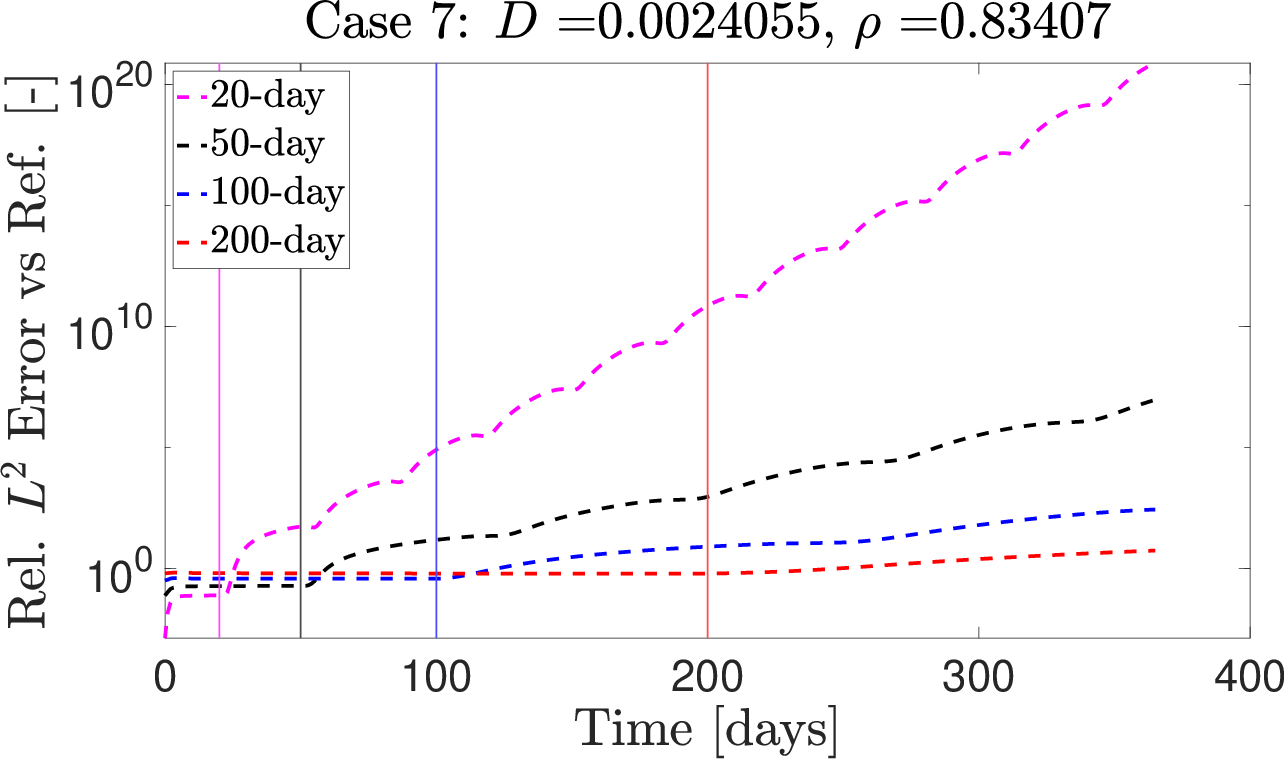}
         \caption{Case $ 7$}
         \label{fig:Case7}
     \end{subfigure}
     \hfill
     \begin{subfigure}[b]{0.31\textwidth}
         \centering
         \includegraphics[width=\textwidth]{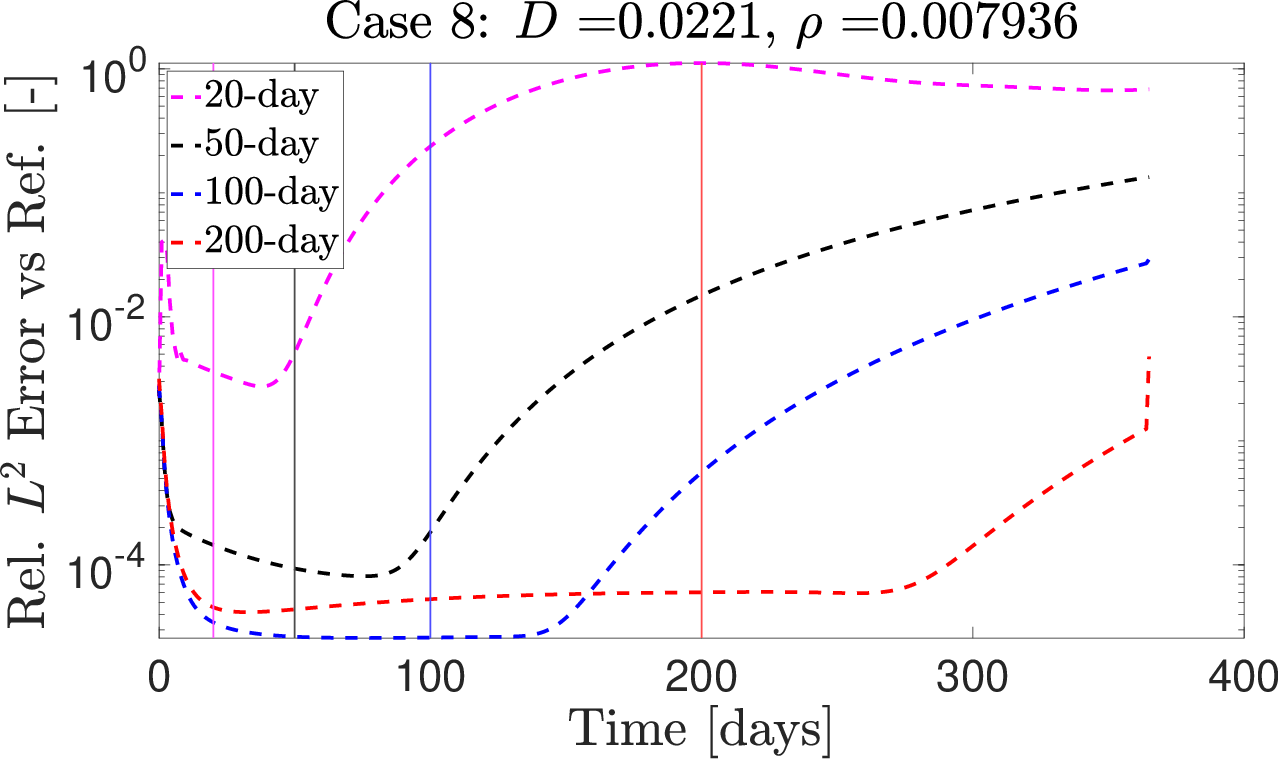}
         \caption{Case $ 8$}
         \label{fig:Case8}
     \end{subfigure}
     \hfill
     \begin{subfigure}[b]{0.31\textwidth}
         \centering
         \includegraphics[width=\textwidth]{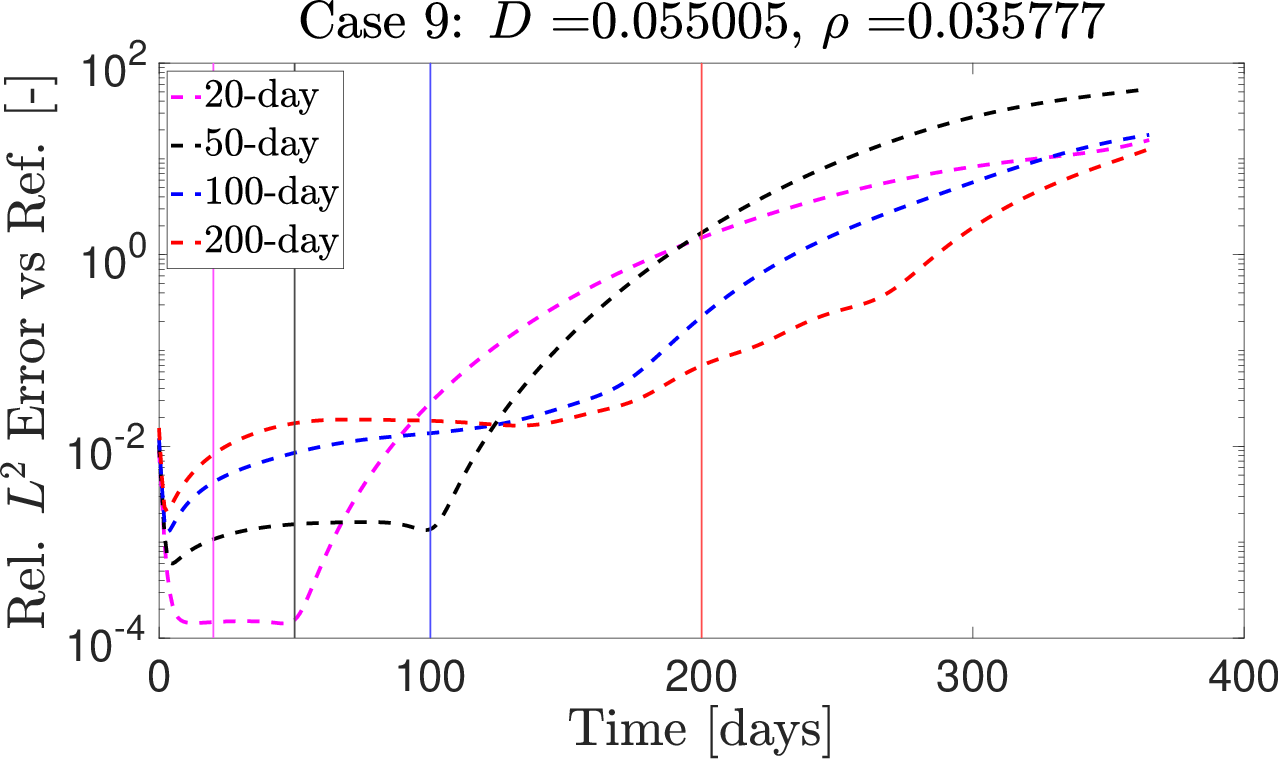}
         \caption{Case $ 9$}
         \label{fig:Case9}
     \end{subfigure}
        \caption{2D simulation study, Cases 7-9. Legend refers to length of the training period.}
        \label{fig:Case7to9}
\end{figure}

\begin{figure}
     \centering
     \begin{subfigure}[b]{0.31\textwidth}
         \centering
         \includegraphics[width=\textwidth]{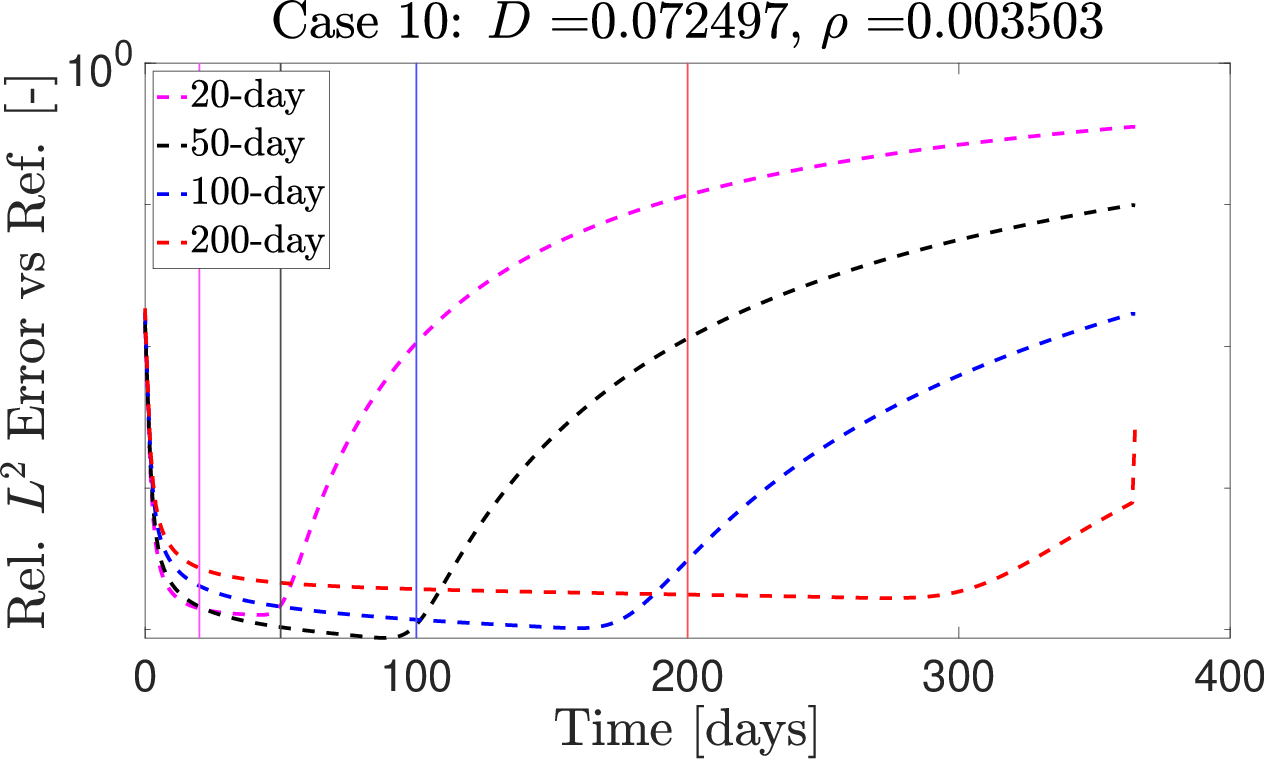}
         \caption{Case $ 10$}
         \label{fig:Case10}
     \end{subfigure}
     \hfill
     \begin{subfigure}[b]{0.31\textwidth}
         \centering
         \includegraphics[width=\textwidth]{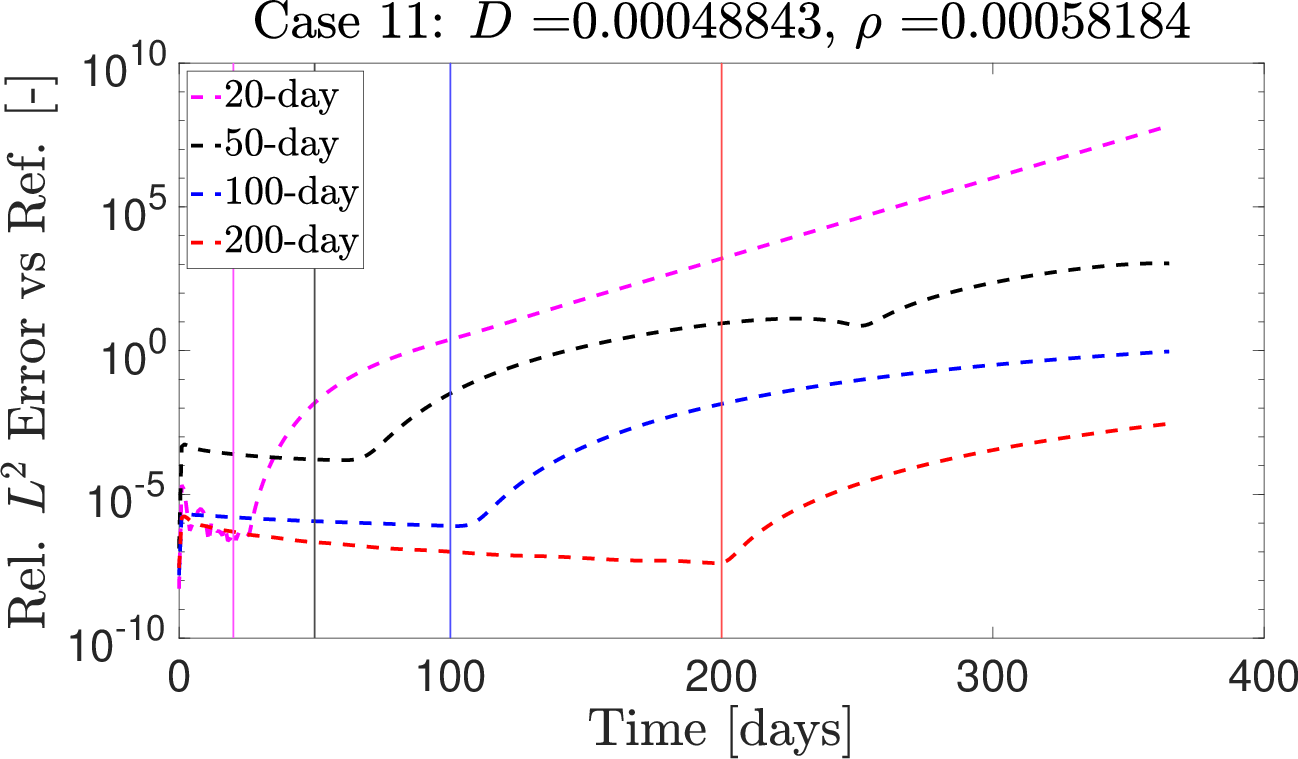}
         \caption{Case $ 11$}
         \label{fig:Case11}
     \end{subfigure}
     \hfill
     \begin{subfigure}[b]{0.31\textwidth}
         \centering
         \includegraphics[width=\textwidth]{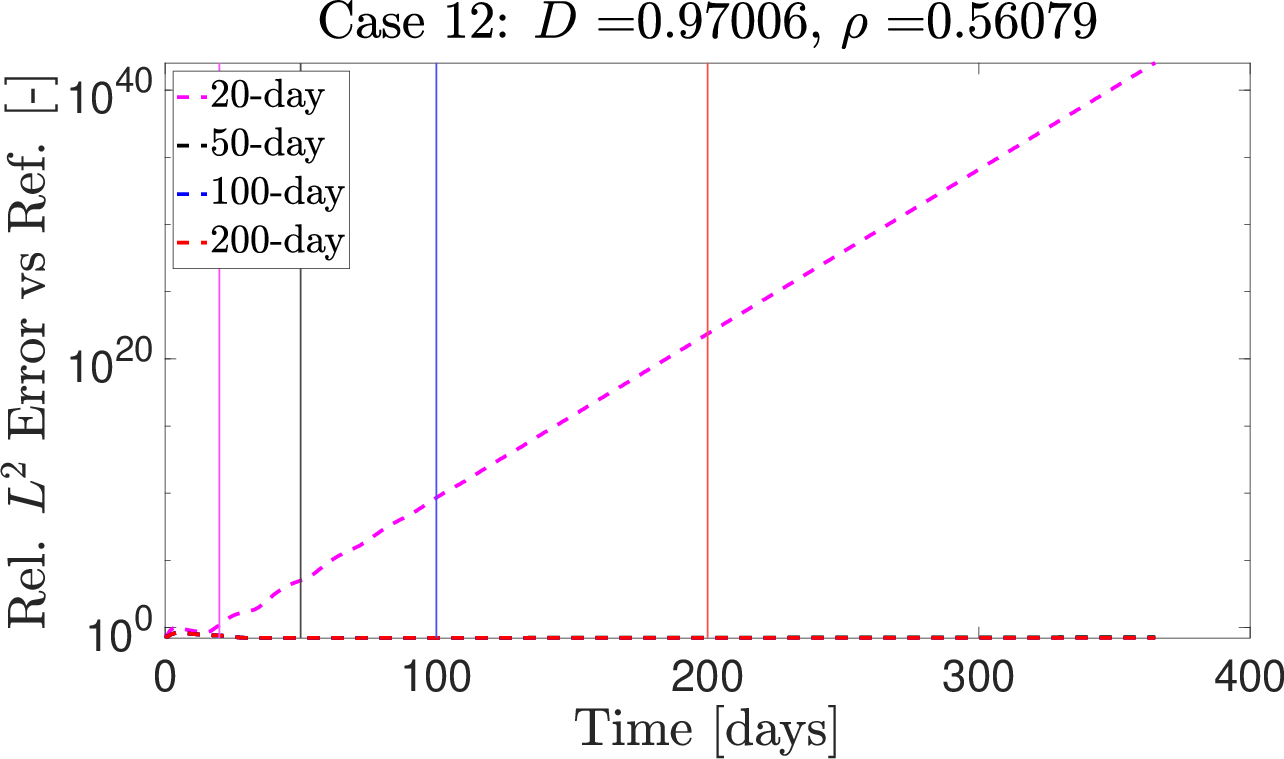}
         \caption{Case $ 12$}
         \label{fig:Case12}
     \end{subfigure}
        \caption{2D simulation study, Cases 10-12. Legend refers to length of the training period.}
        \label{fig:Case10to12}
\end{figure}

\begin{figure}
     \centering
     \begin{subfigure}[b]{0.31\textwidth}
         \centering
         \includegraphics[width=\textwidth]{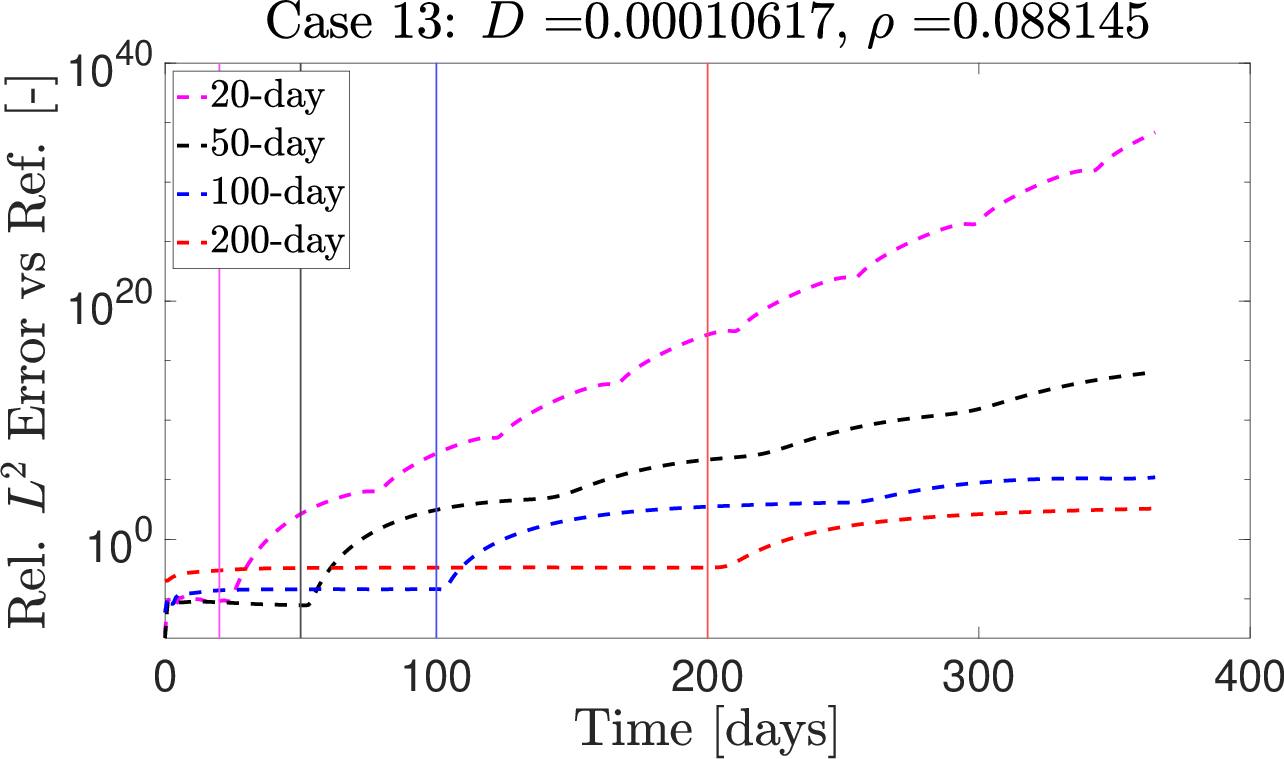}
         \caption{Case $ 13$}
         \label{fig:Case13}
     \end{subfigure}
     \hfill
     \begin{subfigure}[b]{0.31\textwidth}
         \centering
         \includegraphics[width=\textwidth]{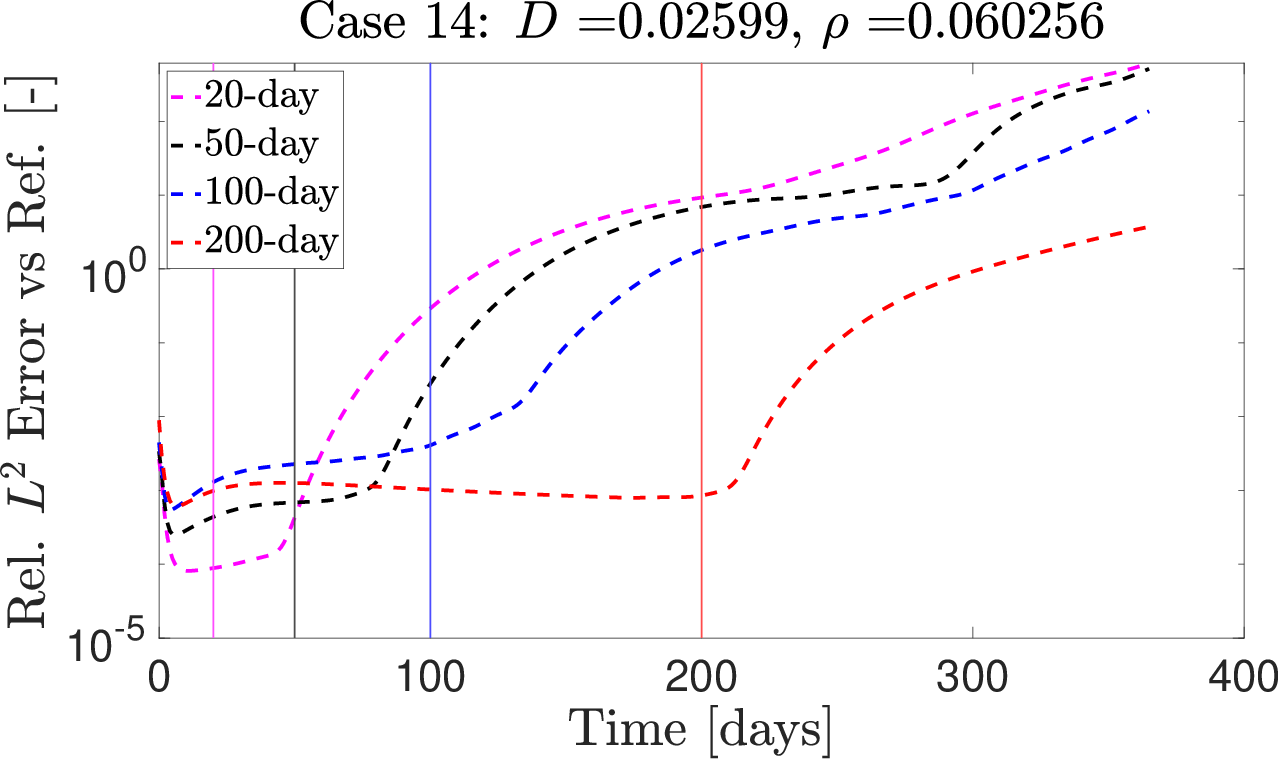}
         \caption{Case $ 14$}
         \label{fig:Case14}
     \end{subfigure}
     \hfill
     \begin{subfigure}[b]{0.31\textwidth}
         \centering
         \includegraphics[width=\textwidth]{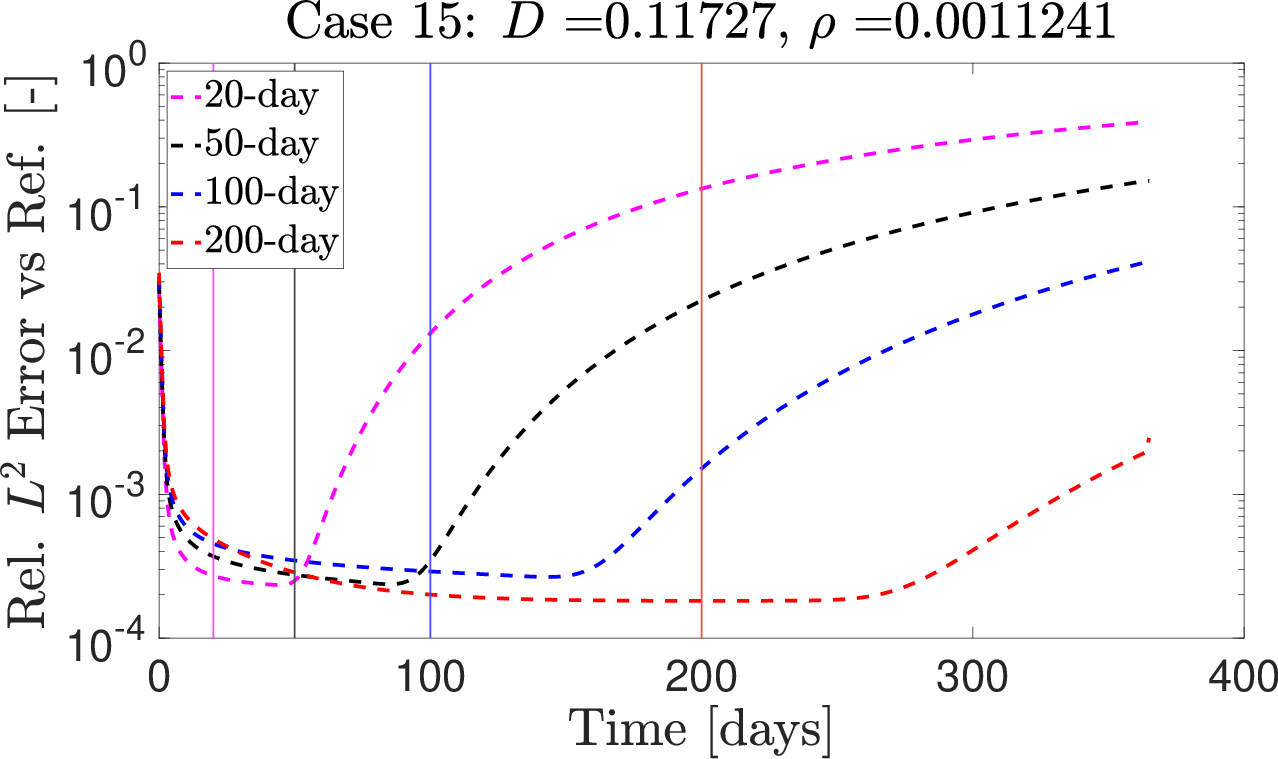}
         \caption{Case $ 15$}
         \label{fig:Case15}
     \end{subfigure}
        \caption{2D simulation study, Cases 13-15. Legend refers to length of the training period.}
        \label{fig:Case13to15}
\end{figure}

\begin{figure}
     \centering
     \begin{subfigure}[b]{0.31\textwidth}
         \centering
         \includegraphics[width=\textwidth]{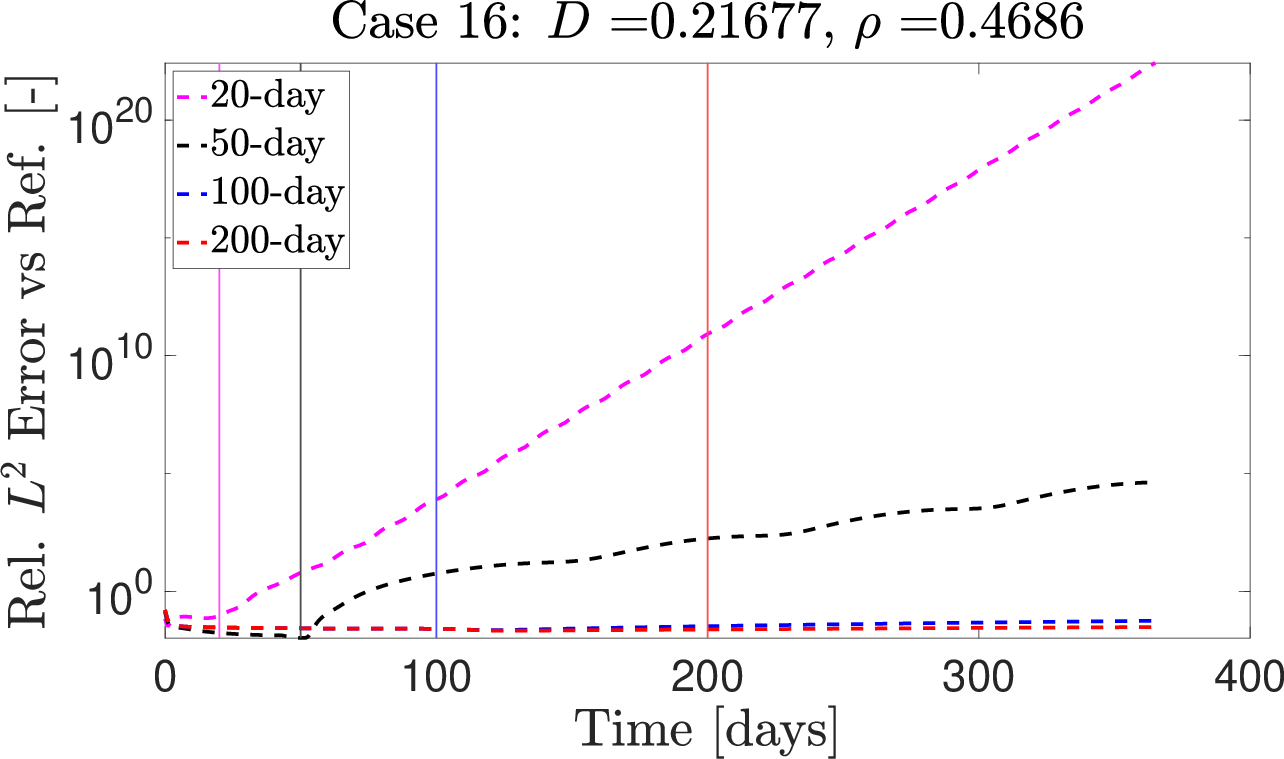}
         \caption{Case $ 16$}
         \label{fig:Case16}
     \end{subfigure}
     \hfill
     \begin{subfigure}[b]{0.31\textwidth}
         \centering
         \includegraphics[width=\textwidth]{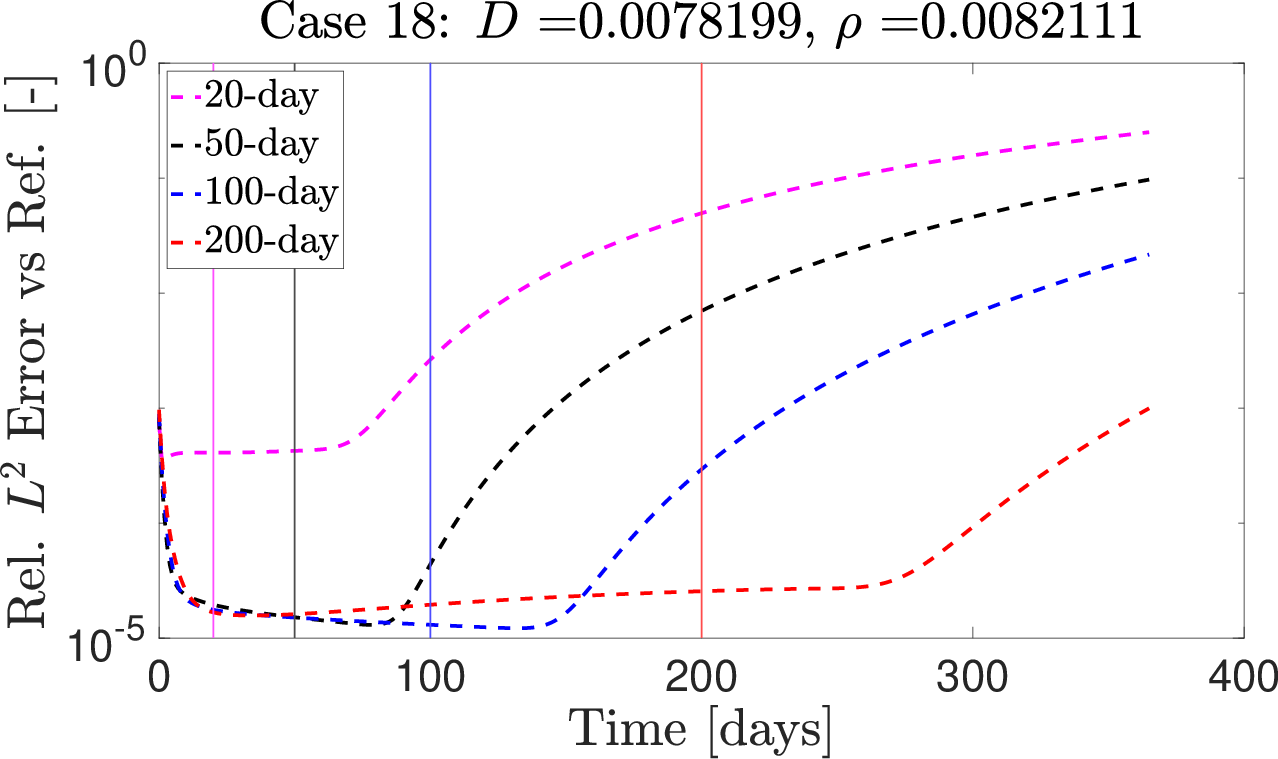}
         \caption{Case $ 18$}
         \label{fig:Case18}
     \end{subfigure}
     \hfill
     \begin{subfigure}[b]{0.31\textwidth}
         \centering
         \includegraphics[width=\textwidth]{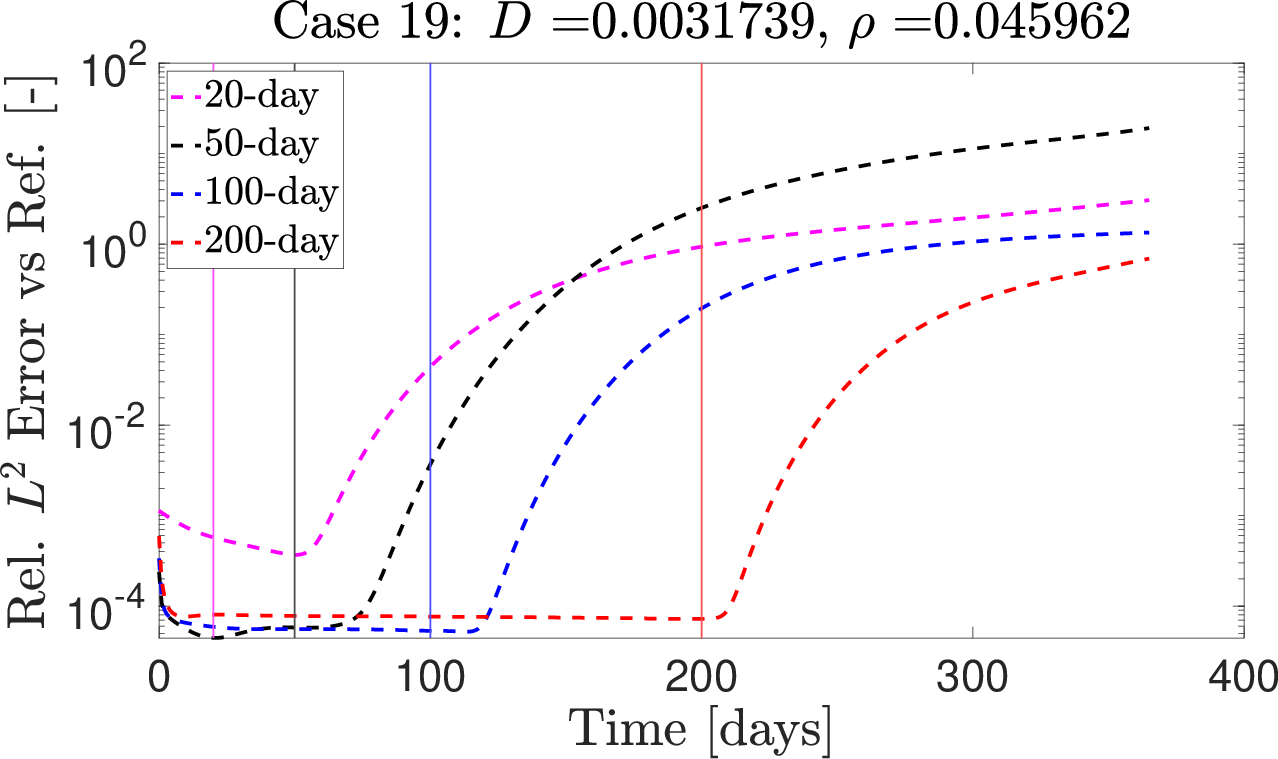}
         \caption{Case $ 19$}
         \label{fig:Case19}
     \end{subfigure}
        \caption{2D simulation study, Cases 16-19 (17 diverged). Legend refers to length of the training period.}
        \label{fig:Case16to19}
\end{figure}

\begin{figure}
     \centering
     \begin{subfigure}[b]{0.31\textwidth}
         \centering
         \includegraphics[width=\textwidth]{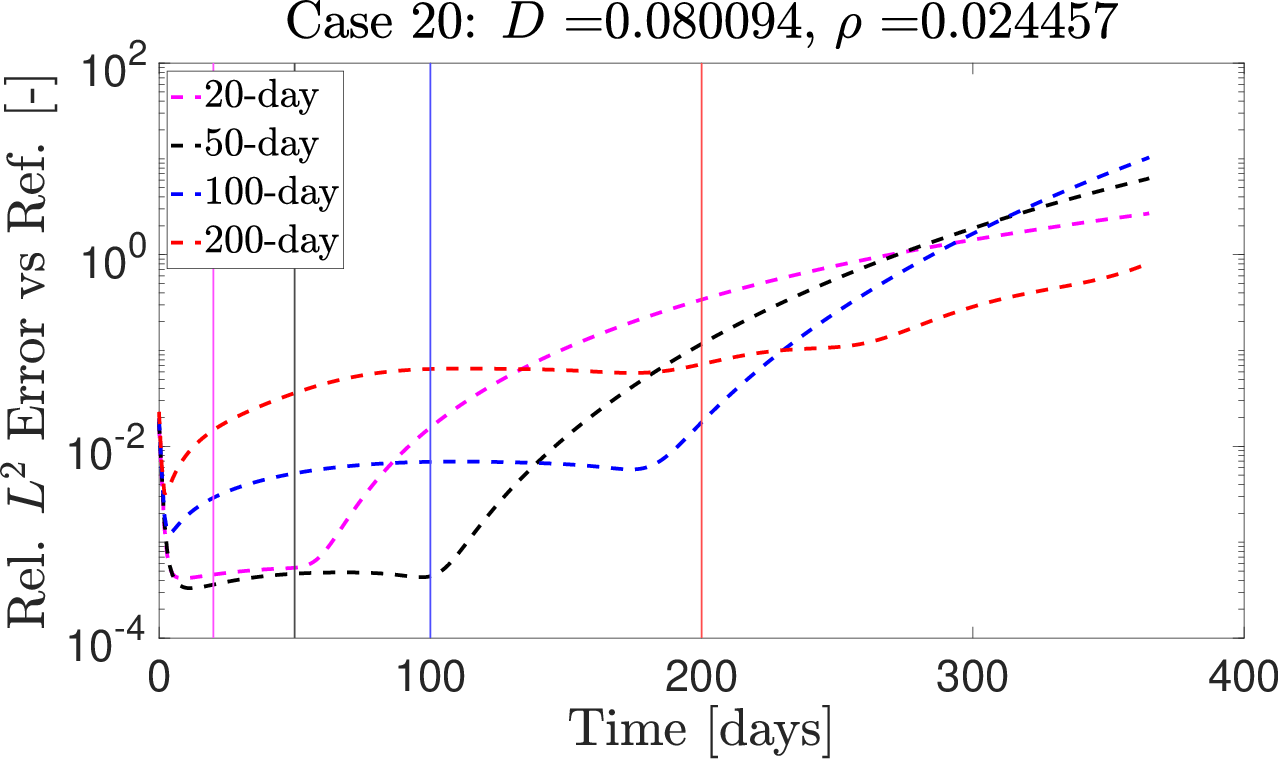}
         \caption{Case $ 20$}
         \label{fig:Case20}
     \end{subfigure}
     \hfill
     \begin{subfigure}[b]{0.31\textwidth}
         \centering
         \includegraphics[width=\textwidth]{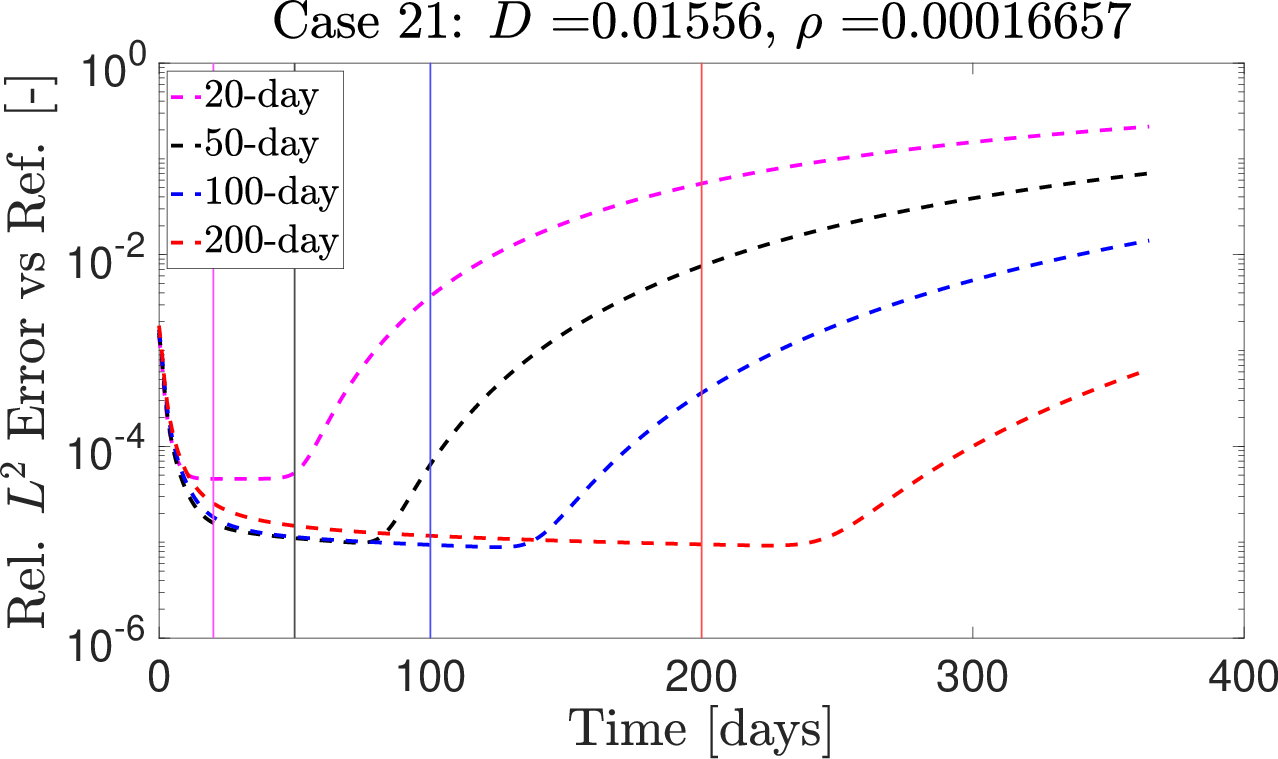}
         \caption{Case $ 21$}
         \label{fig:Case21}
     \end{subfigure}
     \hfill
     \begin{subfigure}[b]{0.31\textwidth}
         \centering
         \includegraphics[width=\textwidth]{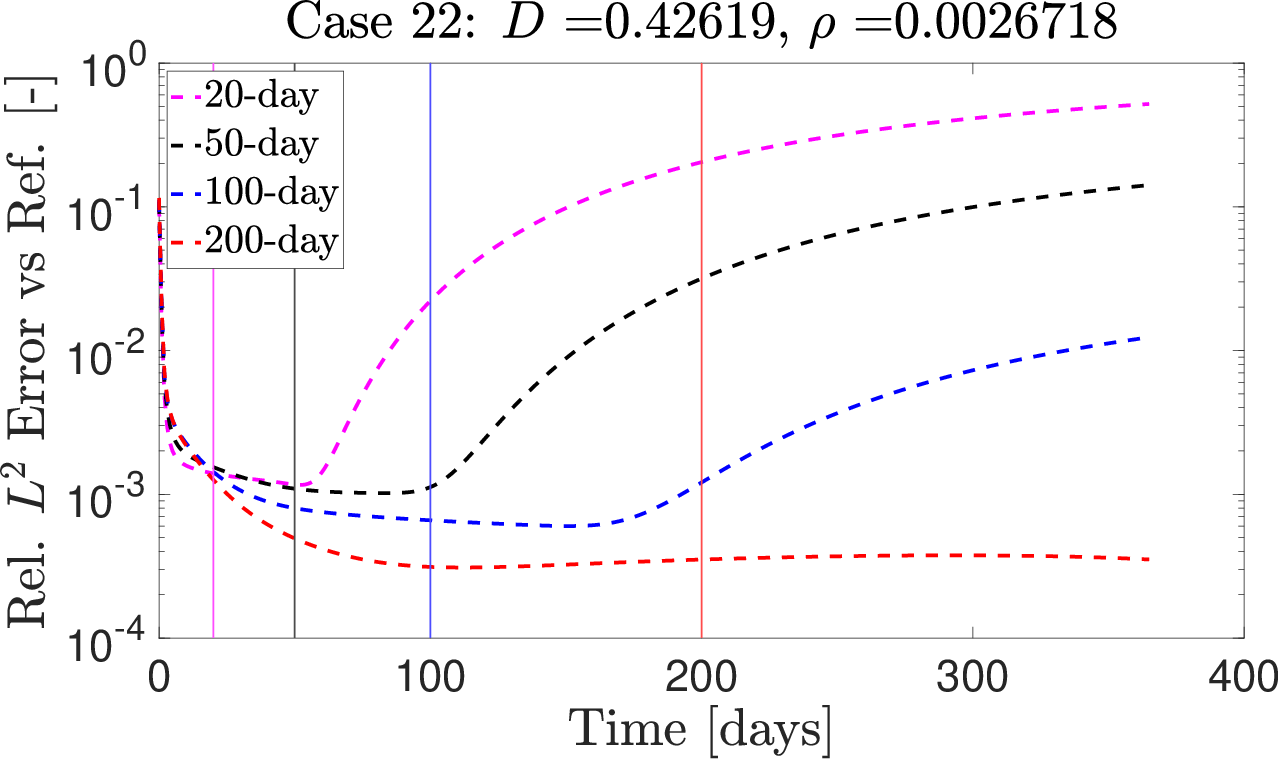}
         \caption{Case $ 22$}
         \label{fig:Case22}
     \end{subfigure}
        \caption{2D simulation study, Cases 20-22. Legend refers to length of the training period.}
        \label{fig:Case20to22}
\end{figure}

\begin{figure}
     \centering
     \begin{subfigure}[b]{0.31\textwidth}
         \centering
         \includegraphics[width=\textwidth]{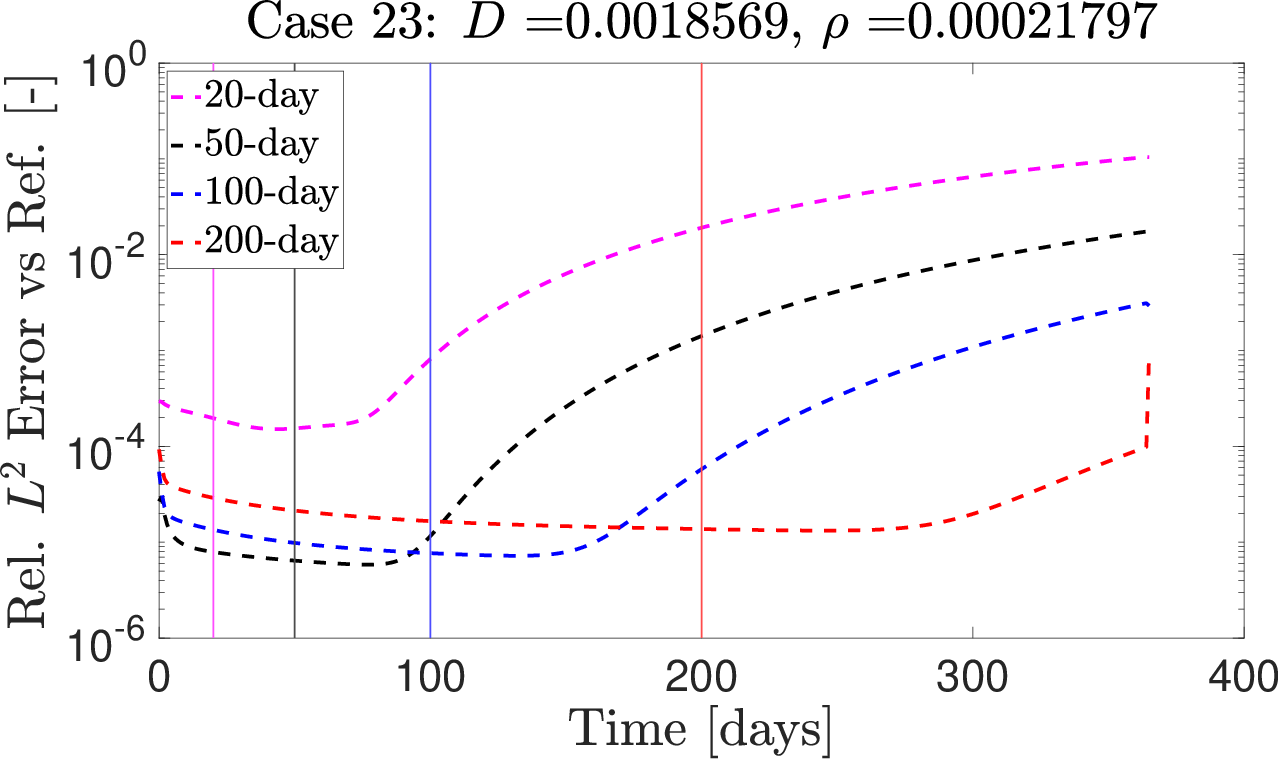}
         \caption{Case $ 23$}
         \label{fig:Case23}
     \end{subfigure}
     \hfill
     \begin{subfigure}[b]{0.31\textwidth}
         \centering
         \includegraphics[width=\textwidth]{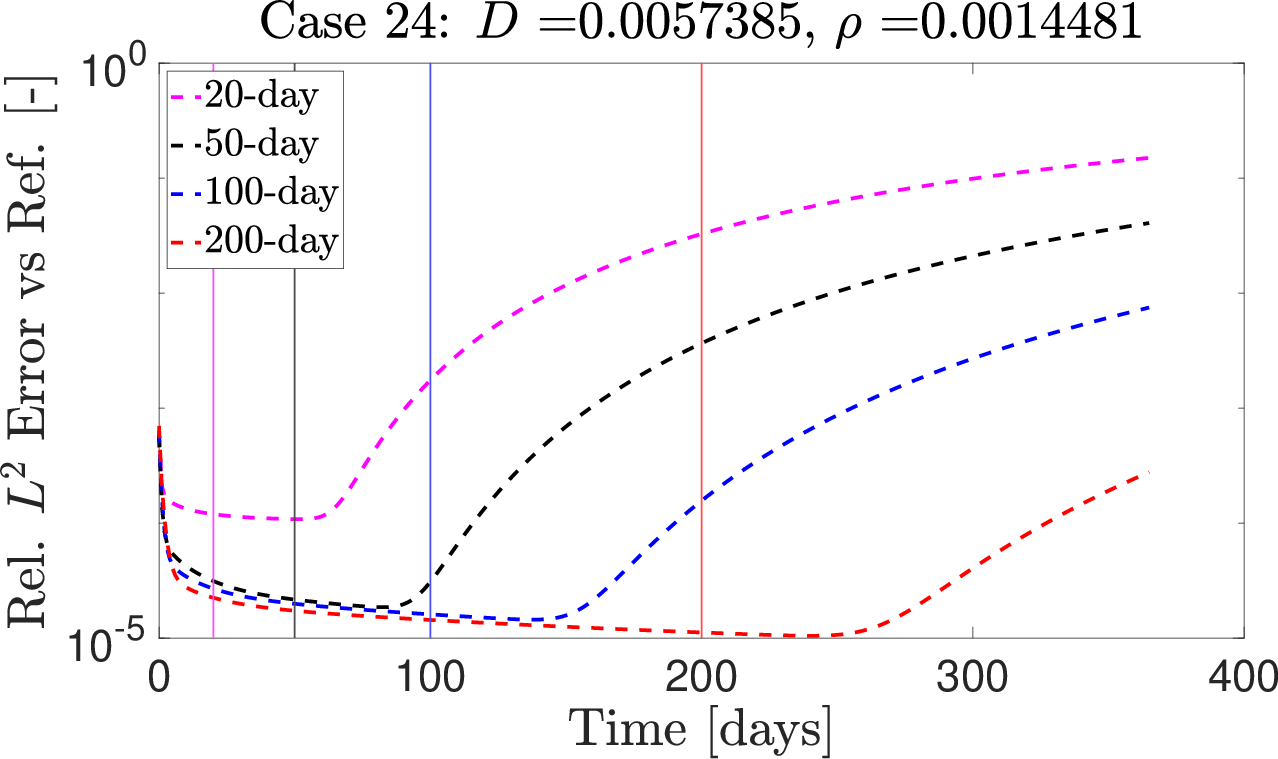}
         \caption{Case $ 24$}
         \label{fig:Case24}
     \end{subfigure}
     \hfill
     \begin{subfigure}[b]{0.31\textwidth}
         \centering
         \includegraphics[width=\textwidth]{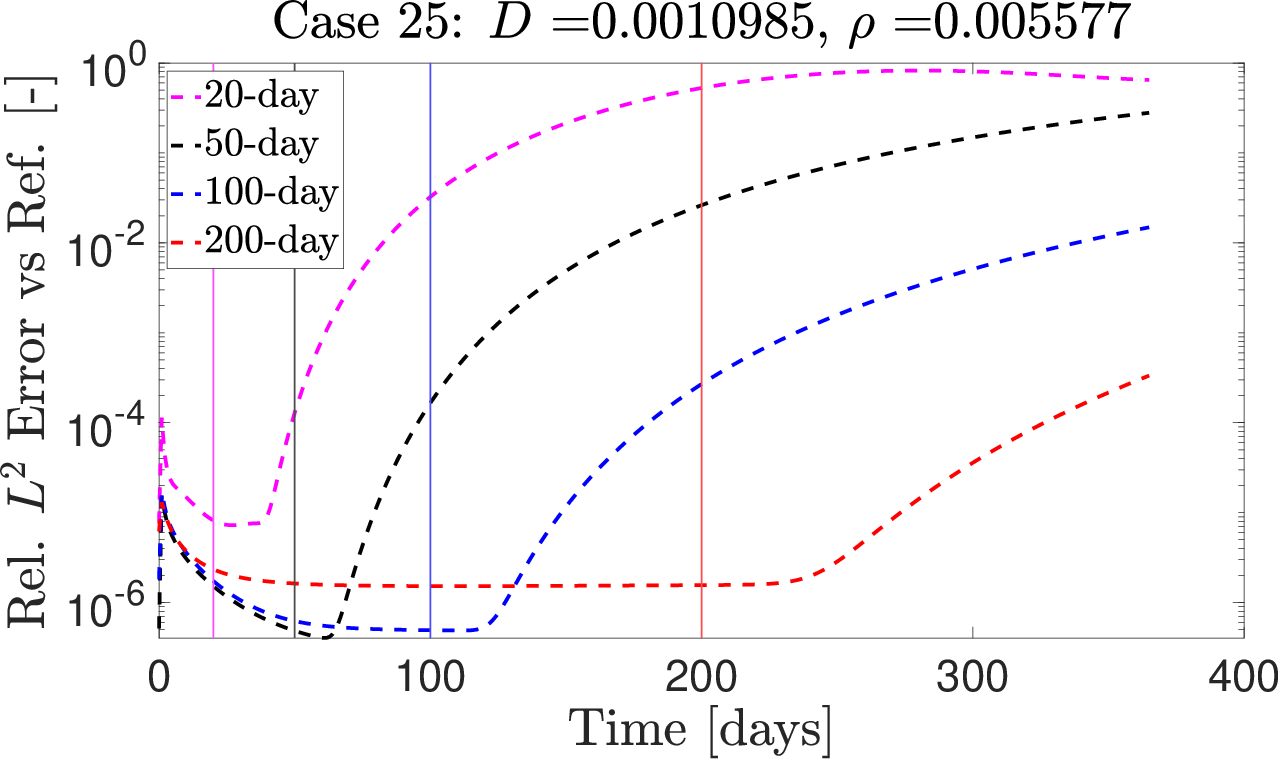}
         \caption{Case $ 25$}
         \label{fig:Case25}
     \end{subfigure}
        \caption{2D simulation study, Cases 23-25. Legend refers to length of the training period.}
        \label{fig:Case23to25}
\end{figure}

\begin{figure}
     \centering
     \begin{subfigure}[b]{0.31\textwidth}
         \centering
         \includegraphics[width=\textwidth]{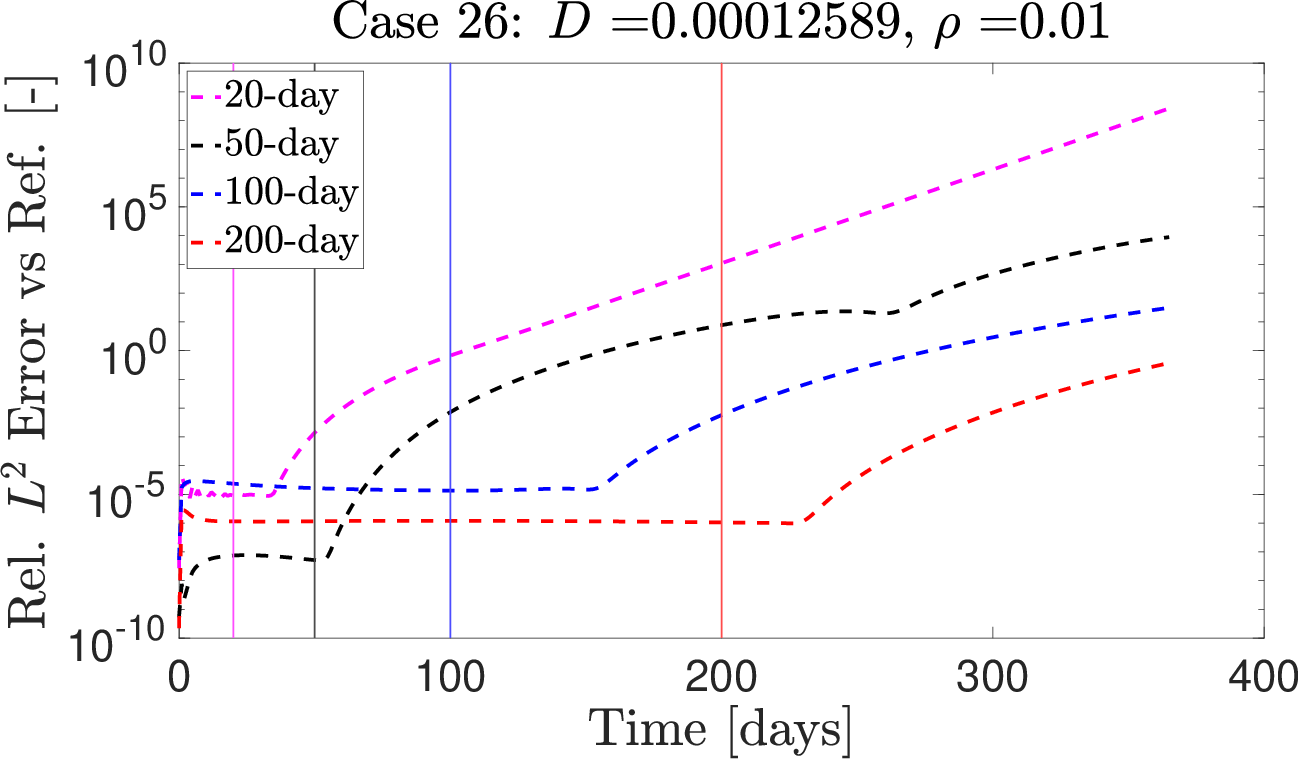}
         \caption{Case $ 26$}
         \label{fig:Case26}
     \end{subfigure}
     \hfill
     \begin{subfigure}[b]{0.31\textwidth}
         \centering
         \includegraphics[width=\textwidth]{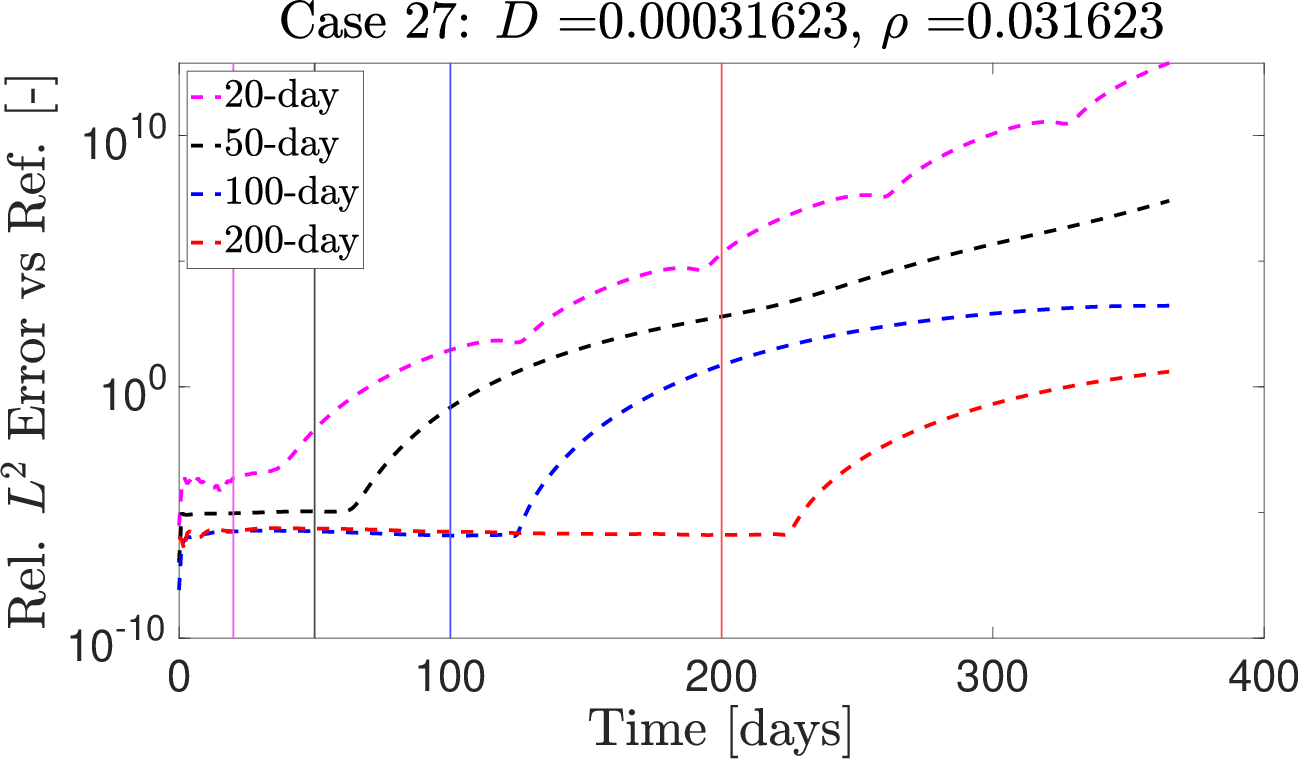}
         \caption{Case $ 27$}
         \label{fig:Case27}
     \end{subfigure}
     \hfill
     \begin{subfigure}[b]{0.31\textwidth}
         \centering
         \includegraphics[width=\textwidth]{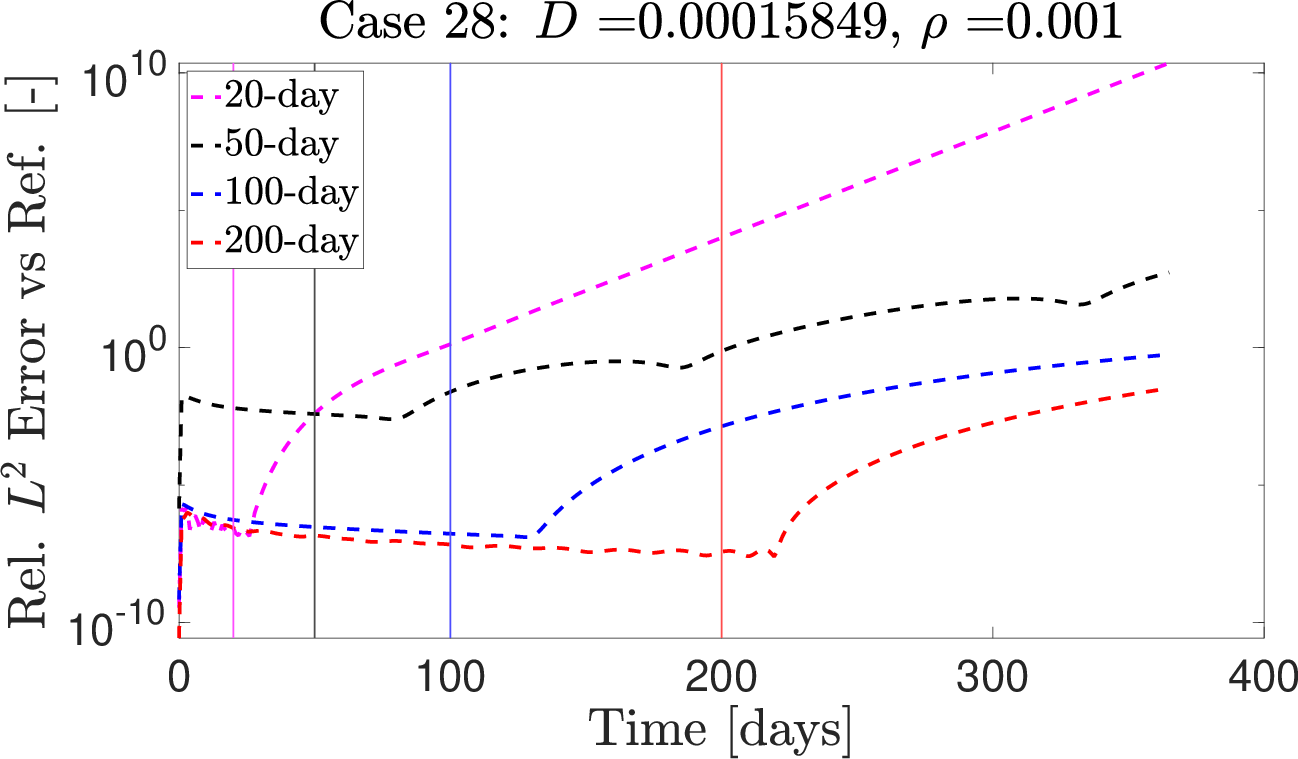}
         \caption{Case $28$}
         \label{fig:Case28}
     \end{subfigure}
        \caption{2D simulation study, Cases 26-28. Legend refers to length of the training period.}
        \label{fig:Case26to28}
\end{figure}

\begin{figure}
     \centering
     \begin{subfigure}[b]{0.31\textwidth}
         \centering
         \includegraphics[width=\textwidth]{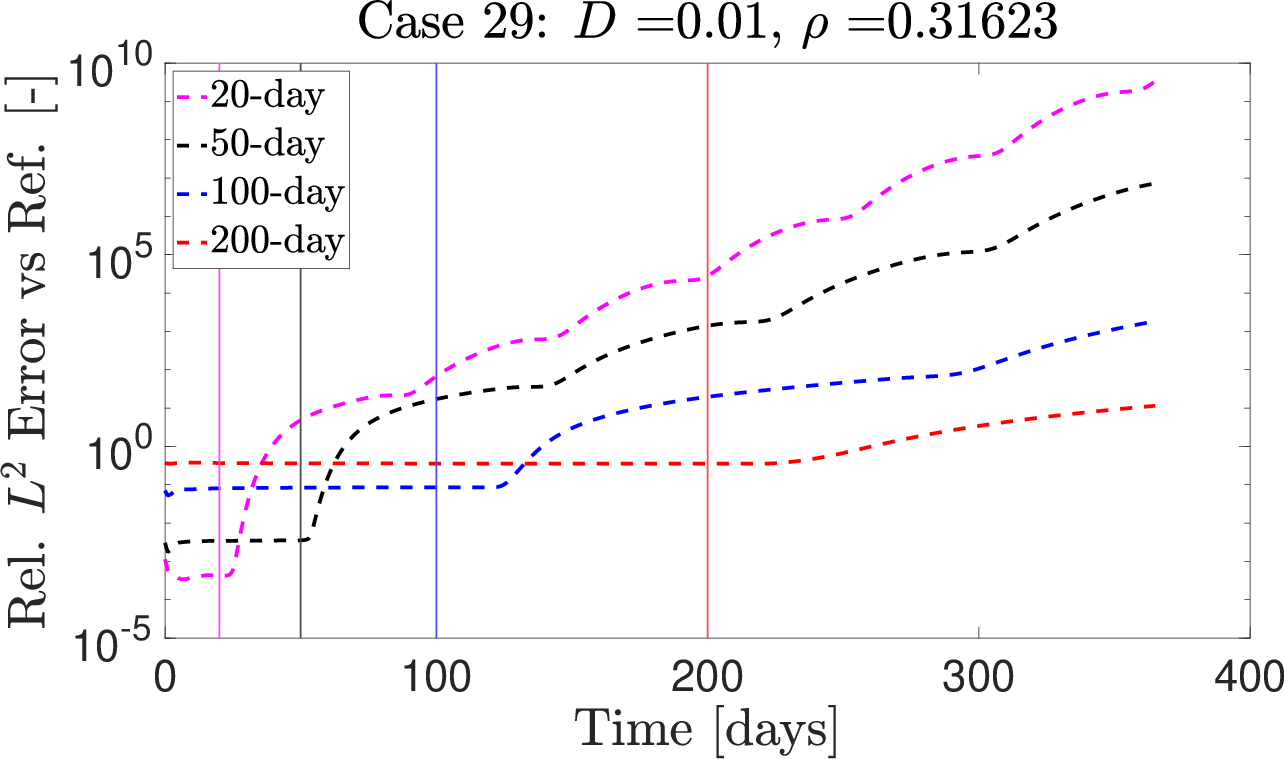}
         \caption{Case $ 29$}
         \label{fig:Case29}
     \end{subfigure}
     \hfill
     \begin{subfigure}[b]{0.31\textwidth}
         \centering
         \includegraphics[width=\textwidth]{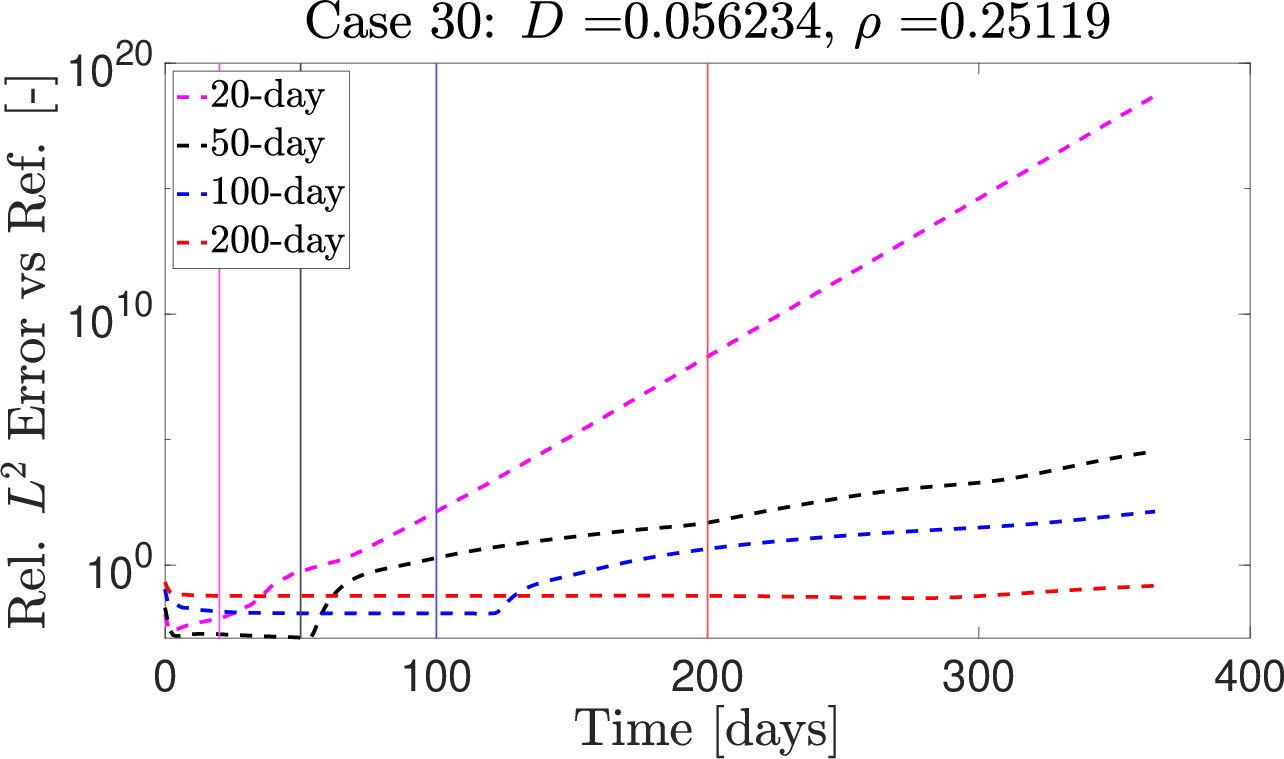}
         \caption{Case $ 30$}
         \label{fig:Case30}
     \end{subfigure}
     \hfill
     \begin{subfigure}[b]{0.31\textwidth}
         \centering
         \includegraphics[width=\textwidth]{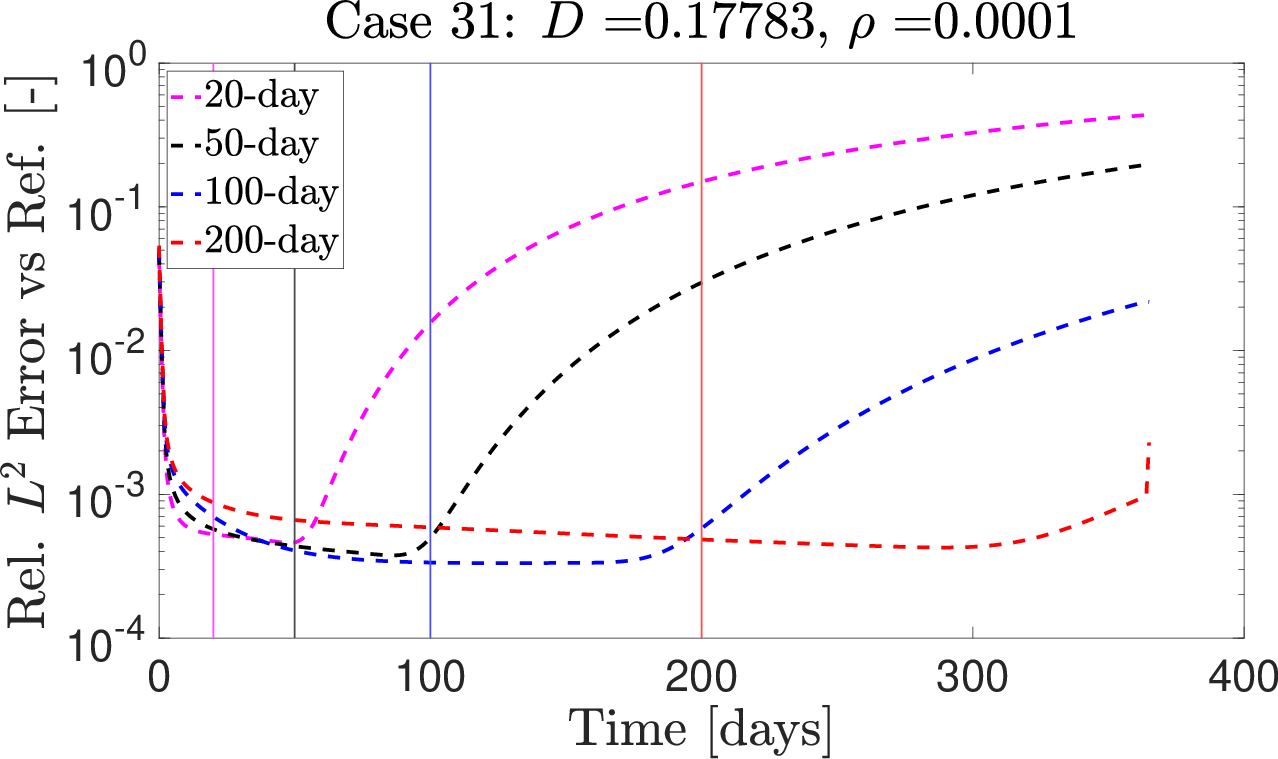}
         \caption{Case $ 31$}
         \label{fig:Case31}
     \end{subfigure}
        \caption{2D simulation study, Cases 29-31. Legend refers to length of the training period.}
        \label{fig:Case29to31}
\end{figure}

\begin{figure}
     \centering
     \begin{subfigure}[b]{0.31\textwidth}
         \centering
         \includegraphics[width=\textwidth]{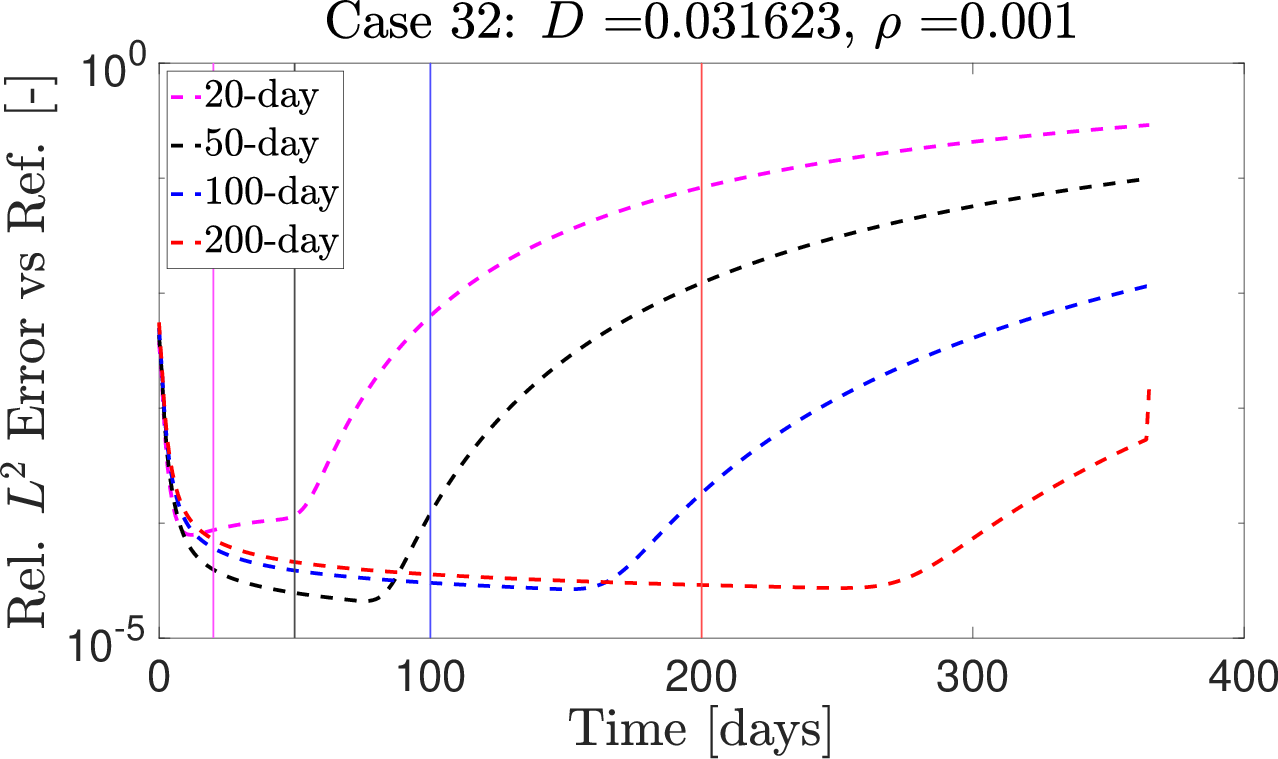}
         \caption{Case $32$}
         \label{fig:Case32}
     \end{subfigure}
     \hspace{.1\textwidth}
     \begin{subfigure}[b]{0.31\textwidth}
         \centering
         \includegraphics[width=\textwidth]{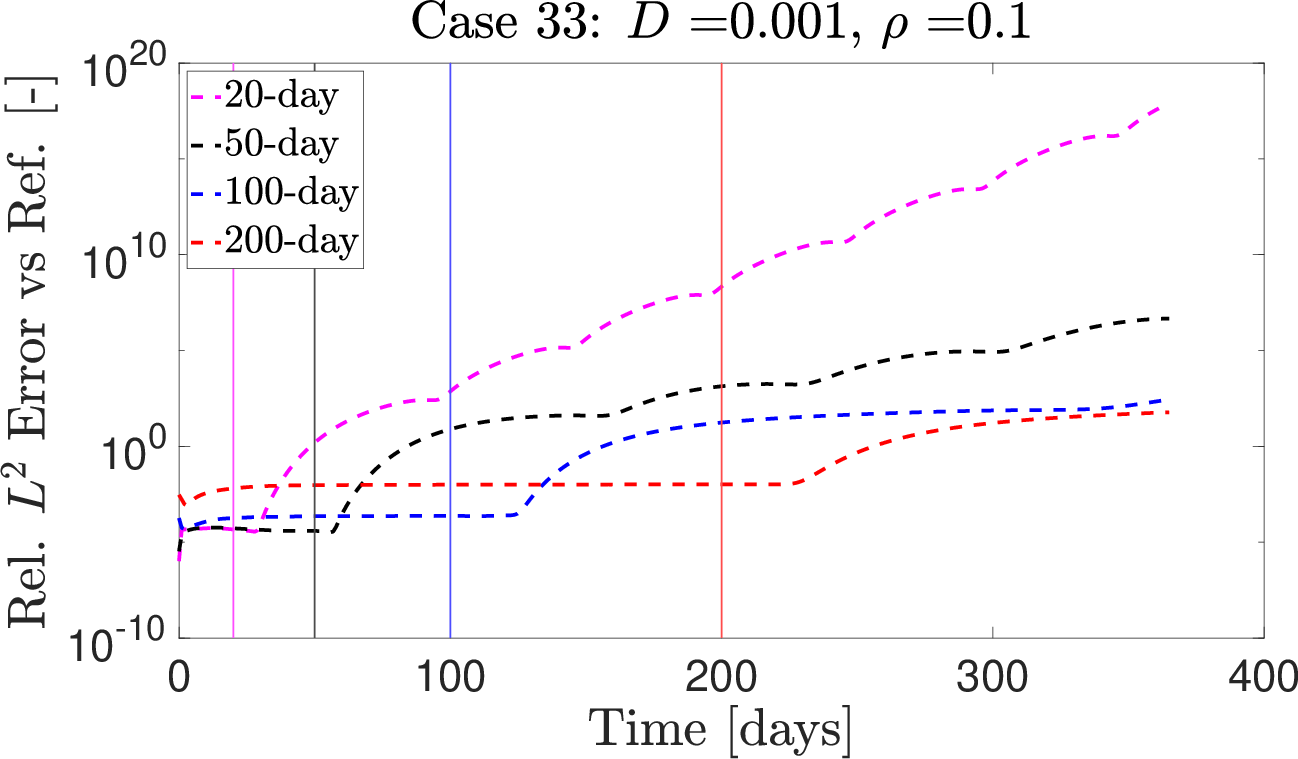}
         \caption{Case $ 33$}
         \label{fig:Case33}
     \end{subfigure}
        \caption{2D simulation study, Cases 32-33. Legend refers to length of the training period.}
        \label{fig:Case32to33}
\end{figure}

\begin{figure}
     \centering
     \begin{subfigure}[b]{0.31\textwidth}
         \centering
         \includegraphics[width=\textwidth]{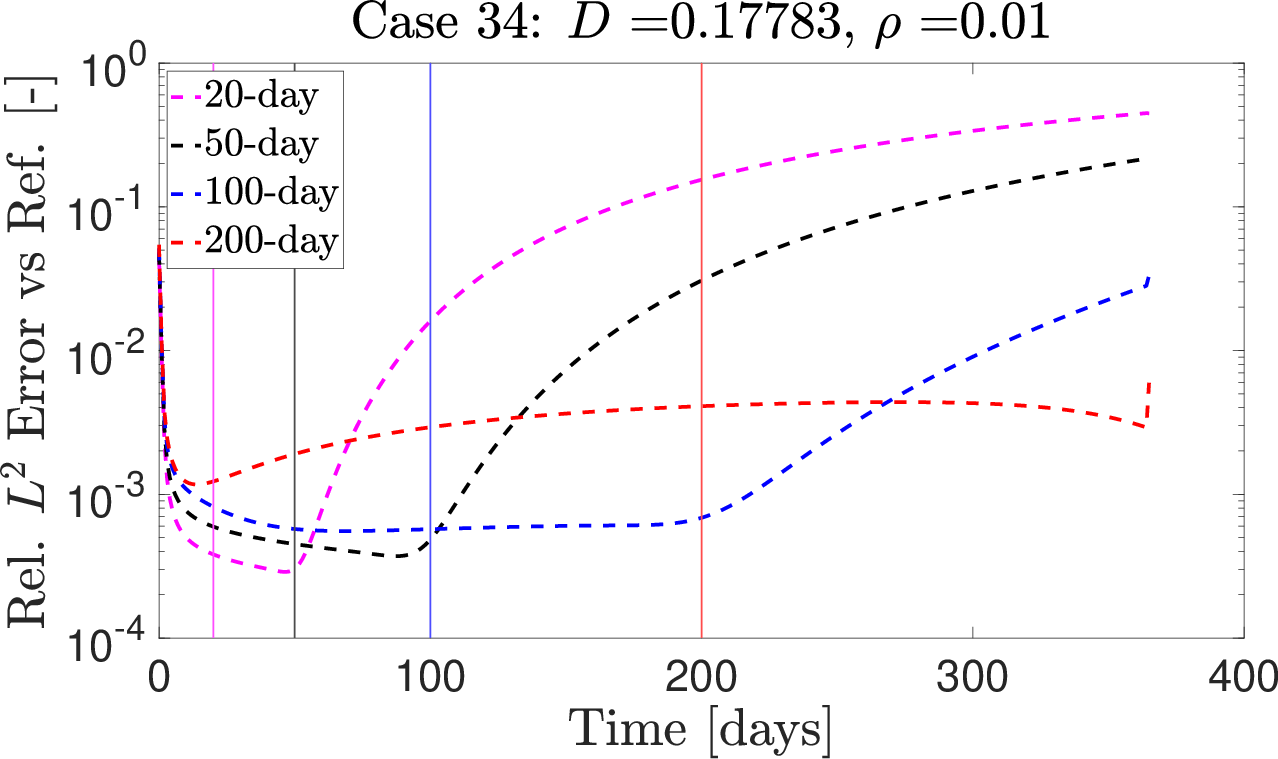}
         \caption{Case $34$}
         \label{fig:Case34}
     \end{subfigure}
     \hspace{.1\textwidth}
     \begin{subfigure}[b]{0.31\textwidth}
         \centering
         \includegraphics[width=\textwidth]{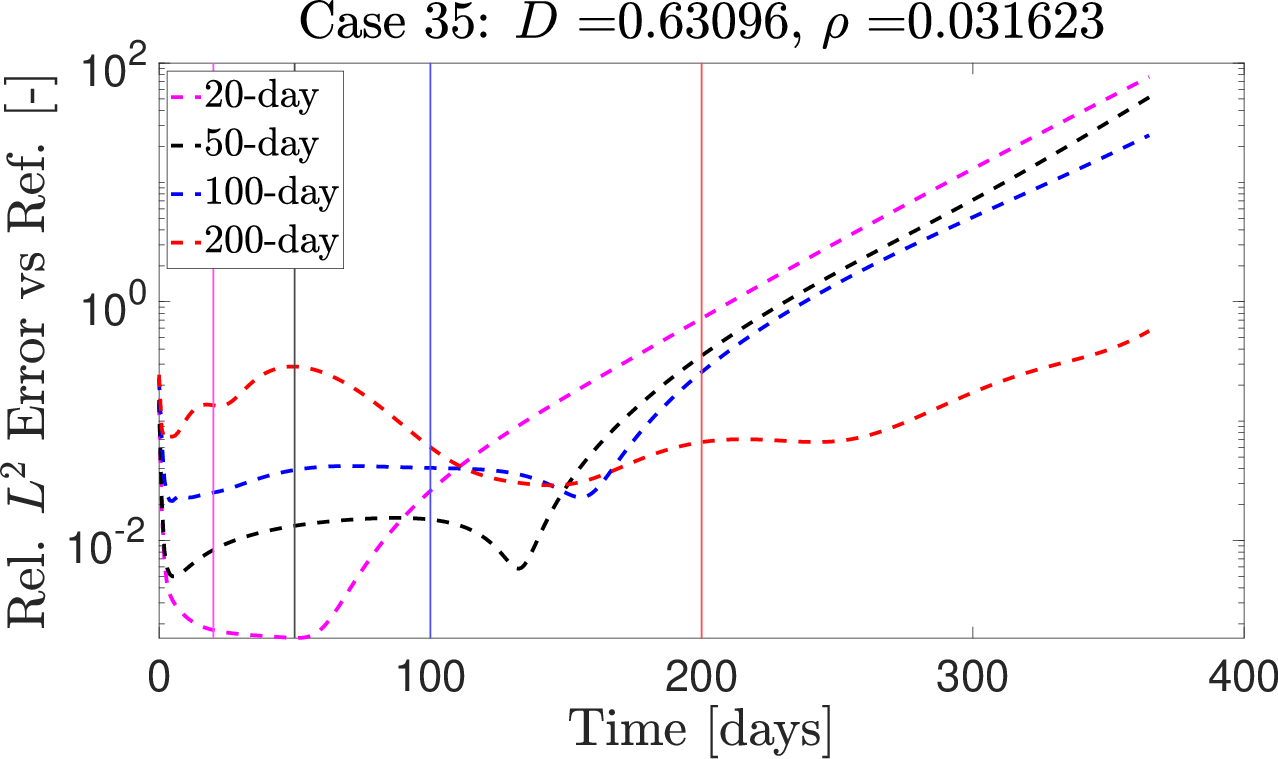}
         \caption{Case $ 35$}
         \label{fig:Case35}
     \end{subfigure}
        \caption{2D simulation study, Cases 34-35. Legend refers to length of the training period.}
        \label{fig:Case34to35}
\end{figure}

 In this appendix we provide the DMD performance plots (see Figures \ref{fig:Case1to3} to \ref{fig:Case34to35}) for all the values of diffusion rate $D$ and proliferation rate $\rho$ considered in the 2D simulation study  (see Tab. \ref{tab:parameter_range}). Values were generated using logarithmic Latin-hypercube sampling.

\begin{table}[ht!]
    \centering
    \begin{tabular}{|c|c|c|}
\hline    
Case	&	$D$	&	$\rho$	\\
\hline
1	&	0.000794	&	0.001835	\\
2	&	0.000499	&	0.315065	\\
3	&	0.249804	&	0.074886	\\
4	&	0.525291	&	0.000504	\\
5	&	0.000111	&	0.000107	\\
6	&	0.004134	&	0.018197	\\
7	&	0.002405	&	0.834065	\\
8	&	0.022100	&	0.007936	\\
9	&	0.055005	&	0.035777	\\
10	&	0.072497	&	0.003503
\\
11	&	0.000488	&	0.000582	\\
12	&	0.970063	&	0.560790	\\
13	&	0.000106	&	0.088145	\\
14	&	0.025990	&	0.060256	\\
15	&	0.117274	&	0.001124	\\
16	&	0.216770	&	0.468598	\\
17	&	0.000277	&	0.235722	\\
18	&	0.007820	&	0.008211	\\
19	&	0.003174	&	0.045962	\\
20	&	0.080094	&	0.024457	\\
21	&	0.015560	&	0.000167	\\
22	&	0.426187	&	0.002672	\\
23	&	0.001857	&	0.000218	\\
24	&	0.005739	&	0.001448	\\
25	&	0.001098	&	0.005577	\\
26	&	0.000126	&	0.010000	\\
27	&	0.000316	&	0.031623	\\
28	&	0.000158	&	0.001000	\\
29	&	0.010000	&	0.316228	\\
30	&	0.056234	&	0.251189	\\
31	&	0.177828	&	0.000100	\\
32	&	0.031623	&	0.001000	\\
33	&	0.001000	&	0.100000	\\
34	&	0.177828	&	0.010000	\\
35	&	0.630957	&	0.031623	\\
\hline
    \end{tabular}
    \caption{Values of diffusion rate $D$ and proliferation rate $\rho$ considered in the 2D simulation study.}
    \label{tab:parameter_range}
\end{table}

%% file: DMDCancer.bbl
\begin{thebibliography}{10}

\bibitem{Agosti2018}
A.~Agosti, C.~Giverso, E.~Faggiano, A.~Stamm, and P.~Ciarletta.
\newblock A personalized mathematical tool for neuro-oncology: A clinical case
  study.
\newblock {\em International Journal of Non-Linear Mechanics}, 107:170--181,
  2018.

\bibitem{Alla2020}
A.~Alla, C.~Balzotti, M.~Briani, and E.~Cristiani.
\newblock {Understanding mass transfer directions via data-driven models with
  application to mobile phone data}.
\newblock {\em SIAM Journal on Applied Dynamical Systems}, 19(2):1372--1391,
  2020.

\bibitem{Alla2017}
A.~Alla and J.~N. Kutz.
\newblock Nonlinear model order reduction via dynamic mode decomposition.
\newblock {\em SIAM Journal on Scientific Computing}, 39(5):B778--B796, 2017.

\bibitem{aronson1978multidimensional}
D.~G. Aronson and H.~F. Weinberger.
\newblock Multidimensional nonlinear diffusion arising in population genetics.
\newblock {\em Advances in Mathematics}, 30(1):33--76, 1978.

\bibitem{artiles2008patch}
W.~Artiles, P.~Carvalho, and R.~A. Kraenkel.
\newblock Patch-size and isolation effects in the {F}isher--{K}olmogorov
  equation.
\newblock {\em Journal of Mathematical Biology}, 57(4):521--535, 2008.

\bibitem{AyalaHernandez2021}
L.~E. Ayala-Hern\'{a}ndez, A.~Gallegos, P.~Schucht, M.~Murek,
  L.~P\'{e}rez-Romasanta, J.~Belmonte-Beitia, and V.~M. P\'{e}rez-Garc\'{i}a.
\newblock Optimal combinations of chemotherapy and radiotherapy in low-grade
  gliomas: A mathematical approach.
\newblock {\em Journal of Personalized Medicine}, 11(10):1036, 2021.

\bibitem{Baldock2014}
A.~L. Baldock, S.~Ahn, R.~Rockne, S.~Johnston, M.~Neal, D.~Corwin,
  K.~Clark-Swanson, G.~Sterin, A.~D. Trister, H.~Malone, et~al.
\newblock Patient-specific metrics of invasiveness reveal significant
  prognostic benefit of resection in a predictable subset of gliomas.
\newblock {\em PLoS One}, 9(10):e99057, 2014.

\bibitem{BGVRC2021}
G.~F. Barros, M.~Grave, A.~Viguerie, A.~Reali, and A.~L. Coutinho.
\newblock Dynamic mode decomposition in adaptive mesh refinement and coarsening
  simulations.
\newblock {\em Engineering with Computers}, 2021.

\bibitem{Ramses2021}
G.~F. Barros, M.~Grave, A.~Viguerie, A.~Reali, and A.~L. Coutinho.
\newblock Enhancing dynamic mode decomposition data pipeline.
\newblock {\em RAMSES: Reduced order models, Approximation theory, Machine
  learning; Surrogates, Emulators and Simulators}, 2021.

\bibitem{beneduci2021unifying}
R.~Beneduci, E.~Bilotta, and P.~Pantano.
\newblock A unifying nonlinear probabilistic epidemic model in space and time.
\newblock {\em Scientific Reports}, 11(1):1--11, 2021.

\bibitem{BradyNicholls2020}
R.~Brady-Nicholls, J.~D. Nagy, T.~A. Gerke, T.~Zhang, A.~Z. Wang, J.~Zhang,
  R.~A. Gatenby, and H.~Enderling.
\newblock Prostate-specific antigen dynamics predict individual responses to
  intermittent androgen deprivation.
\newblock {\em Nature Communications}, 11(1):1750, 2020.

\bibitem{Brueningk2021}
S.~C. Br\"{u}ningk, J.~Peacock, C.~J. Whelan, R.~Brady-Nicholls, M.~Y.
  {Hsiang-Hsuan}, S.~Sahebjam, and H.~Enderling.
\newblock Intermittent radiotherapy as alternative treatment for recurrent high
  grade glioma: A modeling study based on longitudinal tumor measurements.
\newblock {\em Scientific Reports}, 11:20219, 2021.

\bibitem{Calmet2020}
H.~Calmet, D.~Pastrana, O.~Lehmkuhl, T.~Yamamoto, Y.~Kobayashi, K.~Tomoda,
  G.~Houzeaux, and M.~V{\'{a}}zquez.
\newblock {Dynamic Mode Decomposition Analysis of High-Fidelity CFD Simulations
  of the Sinus Ventilation}.
\newblock {\em Flow, Turbulence and Combustion}, 105(3):699--713, 2020.

\bibitem{Chen2013}
X.~Chen, R.~M. Summers, and J.~Yao.
\newblock Kidney tumor growth prediction by coupling reaction–diffusion and
  biomechanical model.
\newblock {\em IEEE Transactions on Biomedical Engineering}, 60(1):169--173,
  2013.

\bibitem{Colli2021}
P.~Colli, H.~Gomez, G.~Lorenzo, G.~Marinoschi, A.~Reali, and E.~Rocca.
\newblock Optimal control of cytotoxic and antiangiogenic therapies on prostate
  cancer growth.
\newblock {\em Mathematical Models and Methods in Applied Sciences},
  31(07):1419--1468, 2021.

\bibitem{el2019revisiting}
M.~El-Hachem, S.~W. McCue, W.~Jin, Y.~Du, and M.~J. Simpson.
\newblock Revisiting the {F}isher--{K}olmogorov--{P}etrovsky--{P}iskunov
  equation to interpret the spreading--extinction dichotomy.
\newblock {\em Proceedings of the Royal Society A}, 475(2229):20190378, 2019.

\bibitem{Ferlay2021}
J.~Ferlay, M.~Colombet, I.~Soerjomataram, D.~M. Parkin, M.~Piñeros, A.~Znaor,
  and F.~Bray.
\newblock Cancer statistics for the year 2020: An overview.
\newblock {\em International Journal of Cancer}, 149(4):778--789, 2021.

\bibitem{Fonzi2020}
N.~Fonzi, S.~L. Brunton, and U.~Fasel.
\newblock {Data-driven nonlinear aeroelastic models of morphing wings for
  control: Data-driven nonlinear aeroelastic models}.
\newblock {\em Proceedings of the Royal Society A: Mathematical, Physical and
  Engineering Sciences}, 476(2239), 2020.

\bibitem{GAO2022110907}
Z.~Gao, Y.~Lin, X.~Sun, and X.~Zeng.
\newblock A reduced order method for nonlinear parameterized partial
  differential equations using dynamic mode decomposition coupled with
  k-nearest-neighbors regression.
\newblock {\em Journal of Computational Physics}, 452:110907, 2022.

\bibitem{giometto2014emerging}
A.~Giometto, A.~Rinaldo, F.~Carrara, and F.~Altermatt.
\newblock Emerging predictable features of replicated biological invasion
  fronts.
\newblock {\em Proceedings of the National Academy of Sciences},
  111(1):297--301, 2014.

\bibitem{grave2020new}
M.~Grave, J.~J. Camata, and A.~L. Coutinho.
\newblock A new convected level-set method for gas bubble dynamics.
\newblock {\em Computers \& Fluids}, 209:104667, 2020.

\bibitem{guozhen1982experiments}
Z.~Guozhen.
\newblock Experiments on director waves in nematic liquid crystals.
\newblock {\em Physical Review Letters}, 49(18):1332, 1982.

\bibitem{henry2006geometric}
D.~Henry.
\newblock {\em Geometric theory of semilinear parabolic equations}, volume 840.
\newblock Springer, 2006.

\bibitem{hess2022data}
M.~W. Hess, A.~Quaini, and G.~Rozza.
\newblock A data-driven surrogate modeling approach for time-dependent
  incompressible navier-stokes equations with dynamic mode decomposition and
  manifold interpolation.
\newblock {\em arXiv preprint arXiv:2201.10872}, 2022.

\bibitem{Hormuth2021}
D.~A. Hormuth, K.~A. Al~Feghali, A.~M. Elliott, T.~E. Yankeelov, and C.~Chung.
\newblock Image-based personalization of computational models for predicting
  response of high-grade glioma to chemoradiation.
\newblock {\em Scientific Reports}, 11(1):8520, 2021.

\bibitem{Jamal2015}
M.~Jamal-Hanjani, S.~A. Quezada, J.~Larkin, and C.~Swanton.
\newblock Translational implications of tumor heterogeneity.
\newblock {\em Clinical Cancer Research}, 21(6):1258--1266, 2015.

\bibitem{Jarrett2018}
A.~M. Jarrett, D.~A. Hormuth, S.~L. Barnes, X.~Feng, W.~Huang, and T.~E.
  Yankeelov.
\newblock Incorporating drug delivery into an imaging-driven, mechanics-coupled
  reaction diffusion model for predicting the response of breast cancer to
  neoadjuvant chemotherapy: theory and preliminary clinical results.
\newblock {\em Physics in Medicine {\&} Biology}, 63(10):105015, 2018.

\bibitem{Jarrett2020}
A.~M. Jarrett, D.~A. Hormuth, C.~Wu, A.~S. Kazerouni, D.~A. Ekrut, J.~Virostko,
  A.~G. Sorace, J.~C. DiCarlo, J.~Kowalski, D.~Patt, B.~Goodgame, S.~Avery, and
  T.~E. Yankeelov.
\newblock Evaluating patient-specific neoadjuvant regimens for breast cancer
  via a mathematical model constrained by quantitative magnetic resonance
  imaging data.
\newblock {\em Neoplasia}, 22(12):820--830, 2020.

\bibitem{Karolak2018}
A.~Karolak, D.~A. Markov, L.~J. McCawley, and K.~A. Rejniak.
\newblock Towards personalized computational oncology: from spatial models of
  tumour spheroids, to organoids, to tissues.
\newblock {\em Journal of the Royal Society Interface}, 15(138):20170703, 2018.

\bibitem{Kazerouni2020}
A.~S. Kazerouni, M.~Gadde, A.~Gardner, D.~A. Hormuth, A.~M. Jarrett, K.~E.
  Johnson, E.~A.~F. Lima, G.~Lorenzo, C.~Phillips, A.~Brock, and T.~E.
  Yankeelov.
\newblock Integrating quantitative assays with biologically based mathematical
  modeling for predictive oncology.
\newblock {\em iScience}, 23(12):101807, 2020.

\bibitem{keller2013numerical}
J.~P. Keller, L.~Gerardo-Giorda, and A.~Veneziani.
\newblock Numerical simulation of a susceptible--exposed--infectious
  space-continuous model for the spread of rabies in raccoons across a
  realistic landscape.
\newblock {\em Journal of Biological Dynamics}, 7(sup1):31--46, 2013.

\bibitem{libmesh}
B.~S. Kirk, J.~W. Peterson, R.~H. Stogner, and G.~F. Carey.
\newblock libmesh: a {C}++ library for parallel adaptive mesh
  refinement/coarsening simulations.
\newblock {\em Journal Engineering with Computers}, 22(3):237--254, 2006.

\bibitem{Kutz2016book}
J.~N. Kutz, S.~L. Brunton, B.~W. Brunton, and J.~L. Proctor.
\newblock {\em Dynamic mode decomposition: data-driven modeling of complex
  systems}.
\newblock SIAM, 2016.

\bibitem{Kutz2016}
J.~N. Kutz, X.~Fu, and S.~L. Brunton.
\newblock {Multiresolution dynamic mode decomposition}.
\newblock {\em SIAM Journal on Applied Dynamical Systems}, 15(2):713--735,
  2016.

\bibitem{Lima2016}
E.~Lima, J.~Oden, D.~Hormuth, T.~Yankeelov, and R.~Almeida.
\newblock Selection, calibration, and validation of models of tumor growth.
\newblock {\em Mathematical Models and Methods in Applied Sciences},
  26(12):2341--2368, 2016.

\bibitem{Lipkova2019}
J.~Lipkov\'{a}, P.~Angelikopoulos, S.~Wu, E.~Alberts, B.~Wiestler, C.~Diehl,
  C.~Preibisch, T.~Pyka, S.~E. Combs, P.~Hadjidoukas, K.~Van~Leemput,
  P.~Koumoutsakos, J.~Lowengrub, and B.~Menze.
\newblock Personalized radiotherapy design for glioblastoma: Integrating
  mathematical tumor models, multimodal scans, and {B}ayesian inference.
\newblock {\em IEEE Transactions on Medical Imaging}, 38(8):1875--1884, 2019.

\bibitem{Litwin2017}
M.~S. Litwin and H.-J. Tan.
\newblock {The Diagnosis and Treatment of Prostate Cancer: A Review}.
\newblock {\em JAMA}, 317(24):2532--2542, 2017.

\bibitem{Lorenzo2021}
G.~Lorenzo, D.~A. Hormuth~II, A.~M. Jarrett, E.~A. Lima, S.~Subramanian,
  G.~Biros, J.~Oden, T.~J. Hughes, and T.~Yankeelov.
\newblock Quantitative in vivo imaging to enable tumor forecasting and
  treatment optimization.
\newblock {\em arXiv}, page 2102.12602, 2021.

\bibitem{Lorenzo2019}
G.~Lorenzo, T.~J.~R. Hughes, P.~Dominguez-Frojan, A.~Reali, and H.~Gomez.
\newblock Computer simulations suggest that prostate enlargement due to benign
  prostatic hyperplasia mechanically impedes prostate cancer growth.
\newblock {\em Proceedings of the National Academy of Sciences of the United
  States of America}, 116(4):1152--1161, 2019.

\bibitem{Lorenzo2019a}
G.~Lorenzo, V.~M. P\'{e}rez-Garc\'{i}a, A.~Mari\~{n}o, L.~A.
  P\'{e}rez-Romasanta, A.~Reali, and H.~Gomez.
\newblock Mechanistic modelling of prostate-specific antigen dynamics shows
  potential for personalized prediction of radiation therapy outcome.
\newblock {\em Journal of The Royal Society Interface}, 16(157):20190195, 2019.

\bibitem{Lorenzo2016}
G.~Lorenzo, M.~A. Scott, K.~Tew, T.~J.~R. Hughes, Y.~J. Zhang, L.~Liu,
  G.~Vilanova, and H.~Gomez.
\newblock Tissue-scale, personalized modeling and simulation of prostate cancer
  growth.
\newblock {\em Proceedings of the National Academy of Sciences of the United
  States of America}, 113(48):E7663--E7671, 2016.

\bibitem{Mang2020}
A.~Mang, S.~Bakas, S.~Subramanian, C.~Davatzikos, and G.~Biros.
\newblock Integrated biophysical modeling and image analysis: Application to
  neuro-oncology.
\newblock {\em Annual Review of Biomedical Engineering}, 22(1):309--341, 2020.

\bibitem{Marusyk2010}
A.~Marusyk and K.~Polyak.
\newblock Tumor heterogeneity: Causes and consequences.
\newblock {\em Biochimica et Biophysica Acta (BBA) - Reviews on Cancer},
  1805(1):105--117, 2010.

\bibitem{MurrayII}
J.~Murray.
\newblock {\em {Mathematical Biology II: Spatial Models and Biomedical
  Application}}.
\newblock Springer, 3 edition, 2003.

\bibitem{MurrayI}
J.~Murray.
\newblock {\em {Mathematical Biology I: An Introduction}}.
\newblock Springer, 3 edition, 2007.

\bibitem{Omuro2013}
A.~Omuro and L.~M. DeAngelis.
\newblock {Glioblastoma and Other Malignant Gliomas: A Clinical Review}.
\newblock {\em JAMA}, 310(17):1842--1850, 2013.

\bibitem{Proctor2015}
J.~L. Proctor and P.~A. Eckhoff.
\newblock {Discovering dynamic patterns from infectious disease data using
  dynamic mode decomposition}.
\newblock {\em International Health}, 7(2):139--145, 2015.

\bibitem{Rockne2019}
R.~C. Rockne, A.~Hawkins-Daarud, K.~R. Swanson, J.~P. Sluka, J.~A. Glazier,
  P.~Macklin, D.~A. Hormuth, A.~M. Jarrett, E.~A. B.~F. Lima, J.~T. Oden,
  G.~Biros, T.~E. Yankeelov, K.~Curtius, I.~A. Bakir, D.~Wodarz, N.~Komarova,
  L.~Aparicio, M.~Bordyuh, R.~Rabadan, S.~D. Finley, H.~Enderling, J.~Caudell,
  E.~G. Moros, A.~R.~A. Anderson, R.~A. Gatenby, A.~Kaznatcheev, P.~Jeavons,
  N.~Krishnan, J.~Pelesko, R.~R. Wadhwa, N.~Yoon, D.~Nichol, A.~Marusyk,
  M.~Hinczewski, and J.~G. Scott.
\newblock The 2019 mathematical oncology roadmap.
\newblock {\em Physical Biology}, 16(4):041005, 2019.

\bibitem{rossa2013parallel}
A.~L. Rossa and A.~L. Coutinho.
\newblock Parallel adaptive simulation of gravity currents on the lock-exchange
  problem.
\newblock {\em Computers \& Fluids}, 88:782--794, 2013.

\bibitem{Taira2017}
K.~Taira, S.~L. Brunton, S.~T. Dawson, C.~W. Rowley, T.~Colonius, B.~J. McKeon,
  O.~T. Schmidt, S.~Gordeyev, V.~Theofilis, and L.~S. Ukeiley.
\newblock {Modal analysis of fluid flows: An overview}.
\newblock {\em AIAA Journal}, 55(12):4013--4041, 2017.

\bibitem{Vavourakis2016}
V.~Vavourakis, B.~Eiben, J.~H. Hipwell, N.~R. Williams, M.~Keshtgar, and D.~J.
  Hawkes.
\newblock Multiscale mechano-biological finite element modelling of oncoplastic
  breast surgery—numerical study towards surgical planning and cosmetic
  outcome prediction.
\newblock {\em PLOS One}, 11(7):e0159766, 2016.

\bibitem{viguerie2022coupled}
A.~Viguerie, G.~F. Barros, M.~Grave, A.~Reali, and A.~L. Coutinho.
\newblock Coupled and uncoupled dynamic mode decomposition in
  multi-compartmental systems with applications to epidemiological and additive
  manufacturing problems.
\newblock {\em Computer Methods in Applied Mechanics and Engineering},
  391:114600, 2022.

\bibitem{Waks2019}
A.~G. Waks and E.~P. Winer.
\newblock {Breast Cancer Treatment: A Review}.
\newblock {\em JAMA}, 321(3):288--300, 2019.

\bibitem{Wang2009}
C.~H. Wang, J.~K. Rockhill, M.~Mrugala, D.~L. Peacock, A.~Lai, K.~Jusenius,
  J.~M. Wardlaw, T.~Cloughesy, A.~M. Spence, R.~Rockne, E.~C. Alvord, and K.~R.
  Swanson.
\newblock Prognostic significance of growth kinetics in newly diagnosed
  glioblastomas revealed by combining serial imaging with a novel
  biomathematical model.
\newblock {\em Cancer Research}, 69(23):9133--9140, 2009.

\bibitem{Wong2017}
K.~C.~L. Wong, R.~M. Summers, E.~Kebebew, and J.~Yao.
\newblock Pancreatic tumor growth prediction with elastic-growth decomposition,
  image-derived motion, and {FDM-FEM} coupling.
\newblock {\em IEEE Transactions on Medical Imaging}, 36(1):111--123, 2017.

\bibitem{Yankeelov2013}
T.~E. Yankeelov, N.~Atuegwu, D.~Hormuth, J.~A. Weis, S.~L. Barnes, M.~I. Miga,
  E.~C. Rericha, and V.~Quaranta.
\newblock Clinically relevant modeling of tumor growth and treatment response.
\newblock {\em Science Translational Medicine}, 5(187):187ps9, 2013.

\bibitem{Zahid2021}
M.~U. Zahid, N.~Mohsin, A.~S. Mohamed, J.~J. Caudell, L.~B. Harrison, C.~D.
  Fuller, E.~G. Moros, and H.~Enderling.
\newblock Forecasting individual patient response to radiation therapy in head
  and neck cancer with a dynamic carrying capacity model.
\newblock {\em International Journal of Radiation Oncology, Biology, Physics},
  111(3):693--704, 2021.

\end{thebibliography}
